\documentclass[jgrga]{AGUTeX}

  \usepackage{graphicx}
  \usepackage{amsmath}
  \usepackage{arydshln}
\authorrunninghead{STERENBORG ET AL}

\titlerunninghead{YOUNG SUN-EARTH INTERACTION}

\authoraddr{M.G. Sterenborg, Department of Earth and Planetary Sciences, Harvard University,  
Cambridge, MA 02138, USA}

\begin{document}

%
%

\title{Modeling the Young Sun's Solar Wind and its Interaction with Earth's Paleomagnetosphere}
%

%
%


\author{M. Glenn Sterenborg}
\affil{Department of Earth and Planetary Sciences,
Harvard University, Cambridge, Massachusetts, USA}

\author{O. Cohen}
\affil{Harvard Smithsonian Center for Astrophysics,
Cambridge, Massachusetts, USA}

\author{J.J. Drake}
\affil{Harvard Smithsonian Center for Astrophysics,
Cambridge, Massachusetts, USA}

\author{T.I. Gombosi}
\affil{Department of Atmospheric, Oceanic, and Space Sciences, University of Michigan, Ann Arbor, MI., USA}



%
%
%

%
%


\begin{abstract}
We present a focused parameter study of solar wind - magnetosphere interaction for the young Sun and Earth, $\sim\!3.5$ Ga ago, that relies on magnetohydrodynamic (MHD) simulations for both the solar wind and the magnetosphere. By simulating the quiescent young Sun and its wind we are able to propagate the MHD simulations up to Earth's magnetosphere and obtain a physically realistic solar forcing of it. We assess how sensitive the young solar wind is to changes in the coronal base density, sunspot placement and magnetic field strength, dipole magnetic field strength and the Sun's rotation period. From this analysis we obtain a range of plausible solar wind conditions the paleomagnetosphere may have been subject to.
Scaling relationships from the literature suggest that a young Sun would have had a mass flux different from the present Sun. We evaluate how the mass flux changes with the aforementioned factors and determine the importance of this and several other key solar and magnetospheric variables with respect to their impact on the paleomagnetosphere. We vary the solar wind speed, density, interplanetary magnetic field strength and orientation as well as Earth's dipole magnetic field strength and tilt in a number of steady-state scenarios that are representative of young Sun-Earth interaction. This study is done as a first step of a more comprehensive effort towards understanding the implications of Sun-Earth interaction for planetary atmospheric evolution.


\end{abstract}

%
%

%

\begin{article}

%
%

\section{Introduction}
The shape and size of Earth's magnetosphere is governed by the interaction of the Earth's magnetic field and ionosphere with the solar wind. As such, the solar wind speed and density, as well as the strength and orientation of both the Earth's magnetic field and the interplanetary magnetic field (IMF) play significant roles. Although the Earth's magnetic field at present has significant power in spherical harmonic degrees beyond the dipole, we may consider it, on average, dipolar with its dipole axis coincident with Earth's rotation axis. However, the field may be considered variable on many spatial and temporal scales. The Earth's dipole moment has seen its magnitude vary a great deal over Earth's history. \cite{tarduno2010} have shown that Earth had a magnetic field as early as $\sim\!3.45$ Ga ago. Their measurements indicated a field strength $50 - 70\%$ of the present field. This constrasts with a scaling relationship that suggests that the field strength scales with the rotation period \citep{stevenson2003}, which would suggest a stronger paleomagnetic field due to Earth's much shorter rotation period in earlier times. Paleomagnetic measurements indicate the field has gone through many polarity reversals \citep{mcfadden2000}. Irregularly spaced in time and of different durations, up to several thousand years, the transitions between polarities are thought to be accompanied by a strong reduction of field strength, down to a tenth of its current strength. It is not yet fully understood how the field goes from one polarity to its opposite but paleomagnetic measurements and dynamo models suggest that the tilt of the dipole can experience excursions prior to reversal that can exceed 45$^\circ$ \citep{glatzmaier1995}. The field's morphology is also thought to change during such a transition, going from mainly dipolar to multipolar \citep{gubbins1994}. All things being equal, such long -and short term changes in the magnetic field will have strongly affected the magnetosphere. \\
The Sun has also changed considerably over its lifetime, \citep{gudel2007,guinan2009}. It is assumed that the Sun has seen stages in its development similar to those observed for younger stars analogous to a younger Sun, on the main-sequence \citep[see, e.g.][]{ribas2005, ribas2010,guinan2009}. In keeping with main-sequence evolution, the solar luminosity is assumed to have increased over its lifetime, but the magnetic activity-related X-ray, EUV and UV outputs have decreased owing to wind-driven magnetic rotational braking and the associated decline in rotation-powered dynamo activity \citep[e.g.][]{skumanich1972,guinan2009}. The solar wind is also highly variable, both on short -and long timescales. For a quiescent Sun we can make the distinction between the fast and slow solar wind in terms of low and high density, respectively.  But for solar flares and coronal mass ejections (CME), the solar wind speed and density can be much greater. Over longer timescales, current observational evidence, albeit scant, indicates that main-sequence stellar wind mass flux declines with age \citep{wood2002, wood2005}. This is important because stellar mass loss rates are directly linked to the magnetic braking that gradually slows stellar rotation and quenches dynamo and related magnetic activity.  Another aspect of Sun-Earth interaction that is linked to the solar mass flux is the Faint Young Sun paradox, which could be resolved by the presence of greenhouse gases in Earth's atmosphere, a lower terrestrial albedo, or an increased mass loss rate of the young Sun \citep{sagan1972, guzik1987, rosing2010}.\\

There is considerable topical interest in solar wind-magnetosphere (-ionosphere) interaction and in particular what this entails for Earth's magnetosphere during the early Archean (3.8 to 3.5 Ga), after the Late Heavy Bombardment and until the first appearance of fossil evidence for simple life \citep{kasting1993}. It is during this time that external modification of Earth's atmosphere (e.g. by impacts) became infrequent enough to allow for a long-term steady state in the physical and chemical properties of the atmosphere. Furthermore, this is as far back in Earth's history that we have evidence of the presence of a planetary magnetic field \citep{tarduno2010}. \\
Understanding the magnetospheric response to variations in solar output is a necessary first step in understanding how these variations ultimately affect Earth's atmosphere and its evolution. This field is also of interest in the larger context of interaction between (magnetized) extrasolar planets and their parent stars and its implications for the development of planetary atmospheres.\\

In this study we focus on how the young Sun interacted with Earth's paleomagnetosphere. Characterization of the paleomagnetosphere, in terms of parameters such as magnetopause stand-off distance, flank distance, polar cap size etc. is often carried out by scaling relationships \citep{siscoe1975, saito1978, vogt2001}. More recently, numerial MHD simulations have been employed to build on this work \citep{zieger2006}, in which several scaling relationships were investigated between the Earth's paleomagnetosphere, its dipole moment and the interplanetary magnetic field. \cite{zieger2004} studied the case of an equatorial dipolar paleomagnetosphere in a large parameter space study by varying the strength of Earth's dipole moment, its dipole tilt and the IMF also using MHD simulations. A similar study looked at quadrupolar paleomagnetospheres as representative of multipolar fields during geomagnetic polarity reversals \citep{vogt2004}. \cite{zieger2006b} researched magnetosphere-ionosphere coupling for the paleomagnetosphere in terms of the transpolar potential and associated field aligned currents. It is noteworthy that many of these studies were carried out for constant solar wind speed and density. In related work but focusing on hot Jupiter-sized exoplanets, \cite{griessmeier2004} used relatively simple scalings for the planetary dipole moment, core radius, magnetopause stand-off distance and stellar wind as well as a parametric model of the magnetosphere, to study their magnetospheric and atmospheric evolution.\\ 

Rather than investigating a large parameter space that is expected to include most Sun-Earth interaction scenarios, from weak multipolar magnetic fields during polarity reversals to strong steady axial dipoles, we attempt a more focused approach aimed at simulating most likely scenarios of young Sun-Earth interaction.  We start by numerically modeling the young Sun and obtain a range of likely values for the solar wind speed and density. These then serve as a forcing of an MHD model of Earth's paleomagnetosphere.  Magnetospheric response to solar output is cast in terms of several key observables such as the subsolar magnetopause stand-off distance, the magnetopause flank distances, polar cap extent and the magnetic and plasma pressures inside the magnetosphere. \\

The main contribution of this study to the field of solar wind-magnetosphere interaction comes from our characterization of the young Sun and obtaining physically jusitifiable solar wind scenarios with which to force the Earth's magnetosphere. This paper, which is readily subdivided in several parts, is structured as follows:
In section \ref{sec:approach} we discuss our approach to modeling Sun-Earth interaction using the Space Weather Modeling Framework, in particular how we model the young Sun and Earth's magnetosphere. We carry out a sensitivity analysis to determine how responsive the solar wind is to changes in sunspot placement, sunspot magnetic field strength, dipole magnetic field strength and rotation period in section \ref{sec:sun}, and in the process obtain solar wind solutions most representative of the young Sun. In section \ref{sec:massflux} we assess how the solar mass flux from our models compare with literature estimates. The MHD simulations of Earth's magnetosphere forced by these solar wind solutions are discussed in section \ref{sec:earth}. We close with our conclusions in section \ref{sec:conclusion}.

\section{Approach}
\label{sec:approach}
For our simulation of the young Sun and its influence on Earth's paleomagnetosphere we use several modules of the Space Weather Modeling Framework (SWMF). This is a high-performance flexible framework for physics-based space weather simulations, as well as for various other space physics applications. The SWMF is able to integrate numerical MHD models of the solar corona, solar eruptive events, the inner heliosphere, solar energetic particles, the global magnetosphere, the inner magnetosphere, the Earth's radiation belt, ionosphere electrodynamics, and the upper atmosphere into a single coupled system. The SWMF enables comprehensive simulations that are not possible with the individual physics models. For a full description of the framework we refer the reader to \citep{powell1999} and \citep{toth2005}. The SWMF delivers the capability to assess the entire system from Sun to Earth in a self-consistent manner and allows us to solve for the MHD variables in Earth's magnetosphere having started from the solar corona.\\

Our approach is straightforward in that we start at the Sun and work our way towards the Earth's magnetosphere. We employ observations of the Sun to constrain a solar corona model and modify it to represent the young Sun. We use this model's output to force the Earth's magnetosphere, the model of which is also modified to represent a paleomagnetosphere, and obtain a physically realistic response. To account for the myriad uncertainties that are inherent in this problem, we vary our input parameters over a range of realistic, physics based, values. We assess the response by inspecting such outputs as subsolar magnetopause stand-off distance $R_{mp}$, magnetopause flank distances $R_{dd}$ (dawn-dusk) and $R_{ns}$ (north-south), polar cap latitudinal extent $\theta_{pc}$ and area $A_{pc}$.

\section{Simulating the young Sun}
\label{sec:sun}
\subsection{Theory}
We use the solar corona (SC) model of the SWMF \citep{cohen2007, cohen2008} to simulate the (young) Sun. 
This model is based on the generic MHD BATS-R-US model \citep{powell1999} and is driven by surface magnetic 
field maps which determine the initial potential magnetic field and are used to scale the boundary conditions 
on the solar photosphere. The energy deposition to the solar wind is based on the observed inverse relation between the 
amount of expansion of a particular magnetic flux tube and the terminal speed of the wind flowing along this tube. 
For an initial potential field distribution (which we calculate using harmonic coefficients supplied by the Wilcox Solar Observatory which observes the Sun's photospheric magnetic field), \cite{wang1990} have 
determined a flux tube expansion factor as:
 \begin{equation}
f_s=\left(\frac{R_{\odot}}{R_{ss}}\right)^2\frac{B(R_{\odot})}{B_0(R_{ss})},
\end{equation} 
where $R_{\odot}$ and $B(R_{\odot})$ are the solar radius and the magnitude of the magnetic field 
at the flux tube base surface, respectively. $B(R_{ss})$ is the magnetic field magnitude of the same flux tube at 
a height $R_{ss}$, which is the height of the so-called ``source surface''. The potential field approximation 
assumes by definition that at the source surface, all field lines are fully radial and open; the source surface is 
commonly set at $R_{ss}=2.5R_\odot$.  Based on the above definition of $f_s$, \cite{arge2000} have derived an 
empirical relation between the expansion factor and the observed solar wind speed at $\sim$1AU (essentially 
the terminal speed) $u_{sw}(r\rightarrow\infty,1/f_s)$. This empirical model is known as the Wang-Sheeley-Arge 
(WSA) model.

In the MHD model used here, the energy necessary to power the solar wind is determined using 
the empirical relation above in the following way. Assuming that $u_{sw}$ is known, we can also assume that the total energy 
of the solar wind far from the Sun equals the bulk kinetic energy of the plasma, $u^2_{sw}/2$. 
On the solar surface, the total energy equals the enthalpy of the plasma minus its gravitational energy. 
Assuming the conservation of energy along a particular flux tube, we can relate the two ends using Bernoulli's 
integral, where we have assumed an adiabatic expansion and a zero speed on the solar surface: 
 \begin{equation}
\frac{u^2_{sw}}{2}=\frac{\gamma_0}{\gamma_0-1}\frac{k_bT_0}{m_p}-\frac{GM_\odot}{R_\odot}.
\label{BI}
\end{equation} 
Here $k_b$ is the Boltzmann constant, $m_p$ is the proton mass, $G$ is the gravitational constant, 
$T_0$ is the boundary condition value for the temperature, and $\gamma_0$ is the surface value 
of the ratio of specific heats. Equation~\ref{BI} enables us to specify the value of $\gamma$ at 
the coronal base as a function of the final speed of the wind that originates from this point. 
The value of $\gamma$ is close to unity near the solar surface but is closer to 1.5 
at 1 AU \citep{totten1995}. A gas with $\gamma\approx 1$ can be considered to have a higher internal 
(turbulent) energy than a gas with $\gamma\approx 1.5$. Therefore, this observed change in the value of 
$\gamma$ can be associated with the amount of internal energy that is released to accelerate 
the solar wind \citep[see][for a full description]{roussev2003}. Based on the surface value 
of $\gamma$ (determined by Equation \ref{BI}), we can define a volumetric energy source term, 
$E_\gamma(\mathbf{r},\gamma_0)$, which vanishes as $\gamma\rightarrow 1.5$. 
With $E_{\gamma}$ and setting an inner boundary condition for the density we now self-consistently solve the set of MHD equations 
\begin{eqnarray}
&\frac{\partial \rho}{\partial t}+\nabla\cdot(\rho \mathbf{u})=0,&  \nonumber \\
&\rho \frac{\partial \mathbf{u}}{\partial t}+\nabla\cdot\left(\rho\mathbf{u}\mathbf{u}+pI+\frac{B^2}{2\mu_0}I-\frac{\mathbf{B}\mathbf{B}}{\mu_0}\right) = \rho\mathbf{g},& \nonumber \\
&\frac{\partial }{\partial t}\left(\frac{1}{2}\rho u^2+\frac{1}{\Gamma-1}p+\frac{B^2}{2\mu_0} \right)+ &  \\
& \nabla\cdot\left(\frac{1}{2}\rho u^2\mathbf{u}+\frac{\Gamma}{\Gamma-1}p\mathbf{u}+\frac{(\mathbf{B}\cdot\mathbf{B})\mathbf{u}-\mathbf{B}(\mathbf{B}\cdot\mathbf{u})}{\mu_0}
\right)=\rho(\mathbf{g}\cdot\mathbf{u})+E_\gamma, & \nonumber \\
&\frac{\partial \mathbf{B}}{\partial t}+\nabla\cdot(\mathbf{u}\mathbf{B}-\mathbf{B}\mathbf{u})=0 & \nonumber
\label{MHD}
\end{eqnarray}
which couples the conservation of mass, momentum, and energy laws as well as the magnetic induction law where $\Gamma$ represents a constant $\gamma=1.5$. 
The $\nabla\cdot\mathbf{B}=0$ condition is enforced using the `eight-wave' method 
\citep{powell1999}. We thus obtain a steady-state solution for the solar wind speed, density, pressure and 
magnetic field strength. In doing so we generate a solar wind solution that is consistent with the distribution 
of the Sun's surface magnetic field and is constrained by the empirical WSA solar wind model. 

In all the SC simulations presented here, we use a non-uniform grid in the SC model ($0-24R_\odot$) with 
grid size of $\Delta x \approx 10^{-2}R_\odot$ near the solar surface. The grid is also dynamically 
refined near the coronal and heliospheric current sheet in order to improve the solution. In the Inner 
Heliosphere (IH) module ($17R_\odot-1~AU$) the grid size near Earth is $\Delta x \approx 0.5R_\odot$. 
SC simulations were performed using the PLEIADES system at the NASA Advanced Supercomputing (NAS) Division.

\subsection{Implementation}
For the purposes of modeling the solar wind and corona, the most relevant feature of a young, `active' Sun is the presence of a stronger surface magnetic field than we observe today.  Simple considerations predict that the efficiency of the magnetic dynamo scales with $\tau\Omega$, where $\tau$ is the convective turnover time in the layers
where the dynamo is operating, and $\Omega$ is the angular velocity of these layers \citep[e.g.][]{durney1978}.   The expected trend of increasing magnetic field strength with increasing rotation rate is borne out, at least qualitatively, by magnetic field observations of solar-like stars \citep[e.g.][]{petit2008}, and the behaviour of magnetic proxy indicators such as X-ray, EUV and UV emissions \citep[e.g.][]{guinan2009, pevtsov2003}. \cite{petit2008} have used the Zeeman Doppler Imaging technique to map out the surface magnetic field of Sun-like stars with different rotation velocities. ÊThey find an increase in the mean stellar magnetic field with rotation velocity, from $3.6\pm 1$~G in HD~146233 with a rotation period of 22.7 days to $42\pm 7$~G in HD~73350 with a period of 12.3 days. ÊHD~146233 is then not only very similar to the present-day Sun in rotation period, but also in terms of its global magnetic field. ÊAt longer rotation periods, poloidal fields (mainly dipolar) completely dominate the global field, as we see on the Sun. ÊHowever, \cite{petit2008} find that toroidal components increase with decreasing period such that 50$\%$ of the magnetic flux of HD~73350 resides in toroidal fields. ÊWe use these magnetic field measurements to guide our simulations. In order to mimic the surface magnetic field of the young Sun, we take detailed observations of the present-day Sun and modify the magnetic field strength and surface distribution of these in ways indicated by observations of young solar analogs.

The baseline for each simulation is a high resolution Michelson Doppler Imager (MDI) magnetogram (http://sun.stanford.edu/synop/) for Carrington Rotation (CR) 1958 (taken in January 2000 during solar maximum). During CR1958 the Sun was relatively active with a large number of active regions on the disk, unlike solar minimum conditions during which the disk is clear of sunspots and the solar magnetic field is nearly dipolar. We `modify' solar activity in two ways. First, we decompose the surface magnetic field in terms of its weak component, which represents its dipolar component, and a strong field component, i.e. the sunspots' magnetic field. The threshold value for this decomposition is 10 G. This allows us to investigate the role each plays by independently scaling them.  Secondly, by moving sunspots to higher latitudes we obtain artifical magnetograms with a realistic spot distribution as observed for young stars such as AB Doradus \citep{hussain2002}. Given that the young Sun is assumed to have a faster rotation rate than the present Sun \citep{gudel2007,guinan2009}, we carry out simulations with the present rotation period of 27 days and with a shorter period of 15 days.
\subsection{Sunspot latitude sensitivity analysis}
\begin{figure}
\noindent\includegraphics[width=39pc]{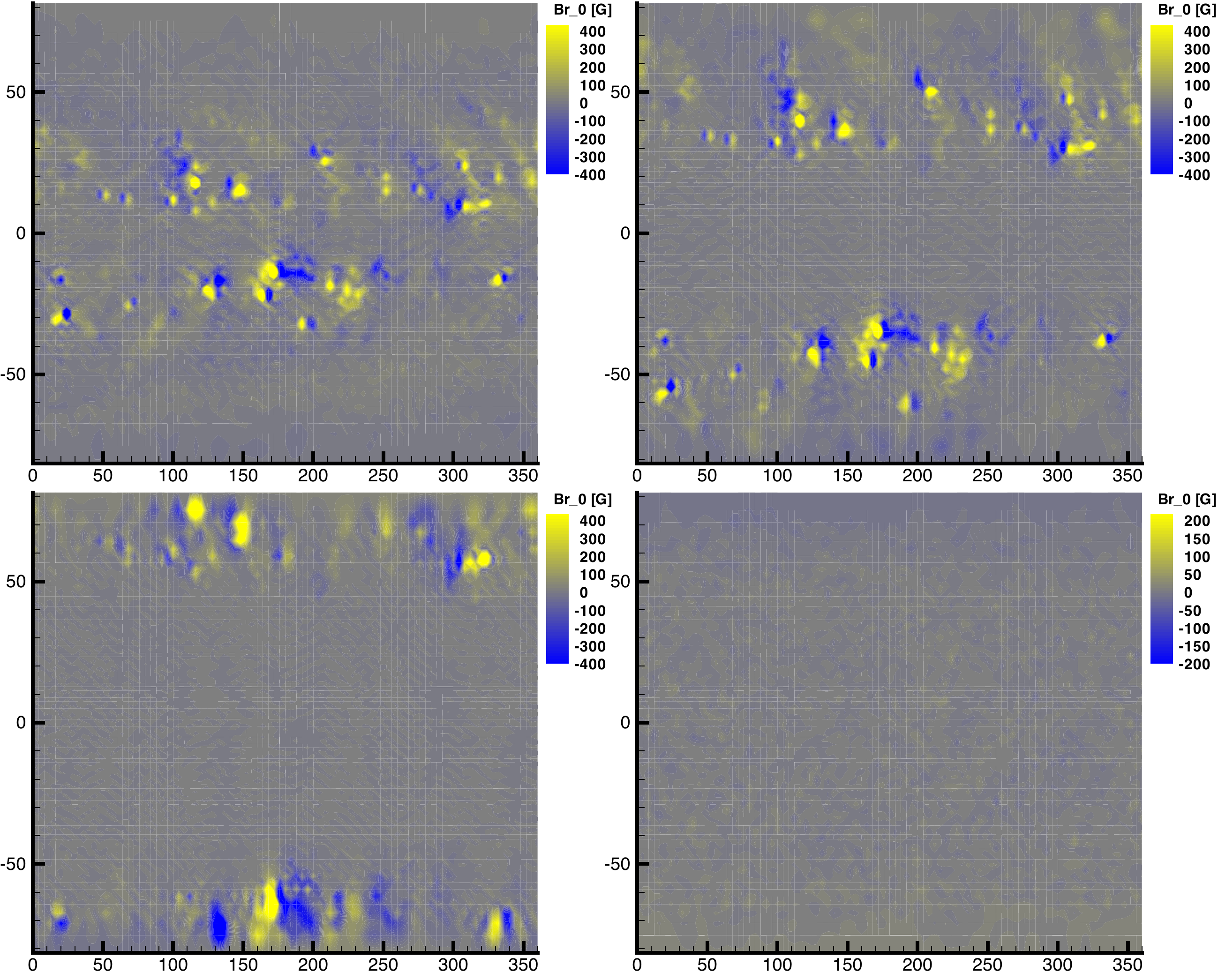}
\caption{Magnetograms used in Figure \ref{fig:scruns2}. MDI magnetograms used in sensitivity analysis of solar wind speed, density, magnetic field strength and mass flux to sunspot placement. Top-left: original magnetogram for a solar maximum (CR1958); top-right: magnetogram for CR1958 with spots placed at $\pm 30^{\circ}$ latitude; bottom-left: magnetogram for CR1958 with spots placed at $\pm 60^{\circ}$ latitude; bottom-right: original magnetogram for a solar minimum (CR2074).}
\label{fig:mgram2}
\end{figure}

\begin{figure}
\noindent\includegraphics[width=39pc]{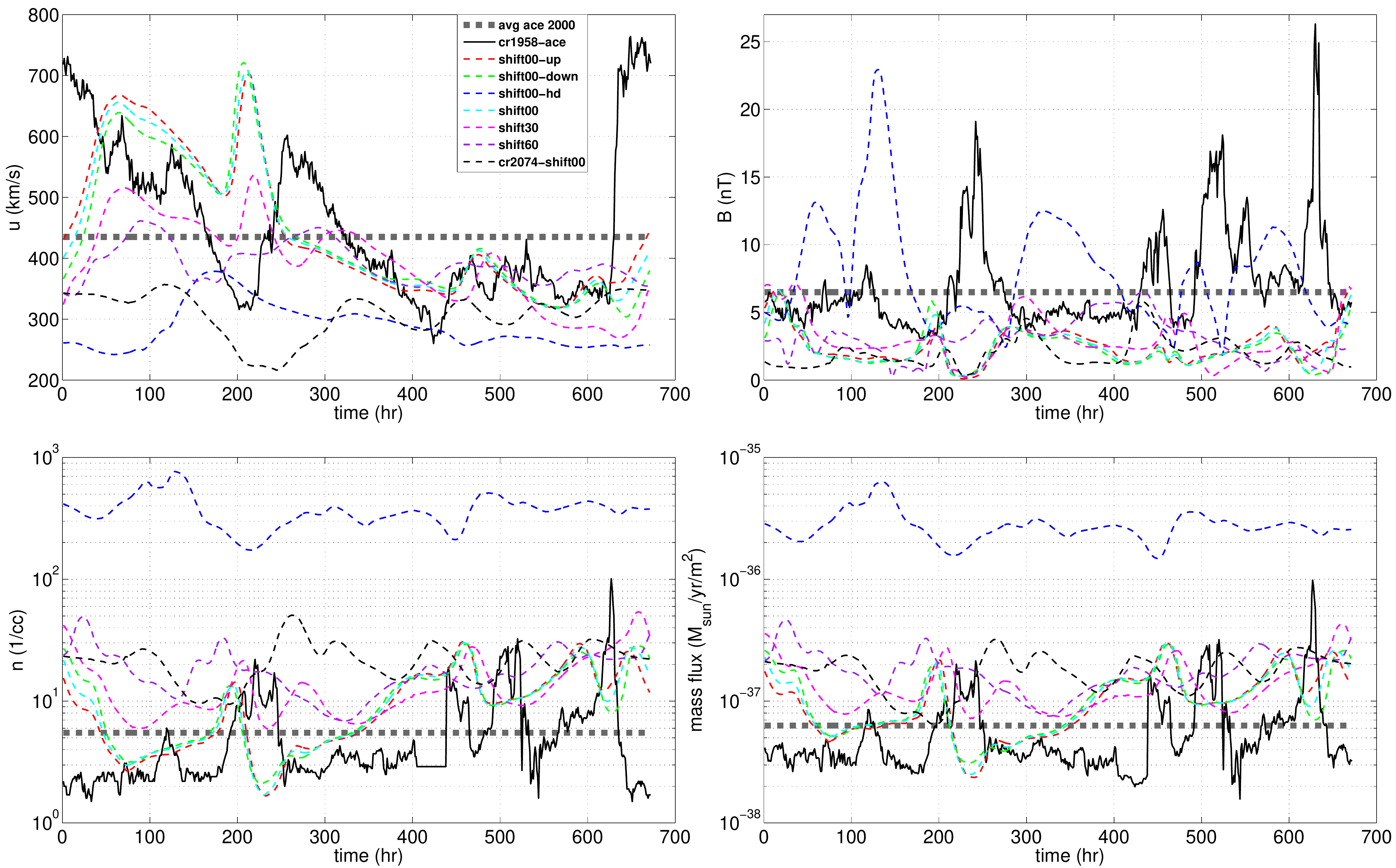}
\caption{Seven SC solar wind solutions at 1 AU along Earth's orbit with a sunspot magnetic field scaling factor of 2, a dipole magnetic field scaling factor of 2  and for a 27 day rotation period. The period covers 28 days. The panels show plasma speed $u$ (top-left), magnetic field strength $B$ (top-right), plasma number density $n$ (bottom-left) and mass flux (bottom-right).
In the legend `shift' refers to which latitude the sunspots have been shifted to. The solution called `up' refers to the solar wind solution evaluated 15$^{\circ}$ above Earth's orbit and `down' below. The `hd' solution refers to a coronal base density of $10^{10}$ cm$^{-3}$ instead of the regular 5$\times10^8$ cm$^{-3}$. We compare the solar wind solutions to ACE data at 1 AU for CR1958 (from January 1, 2000, to January 29, 2000) as well as the average values of ACE data for the year 2000. }
\label{fig:scruns2}
\end{figure}

\begin{figure}
\noindent\includegraphics[width=39pc]{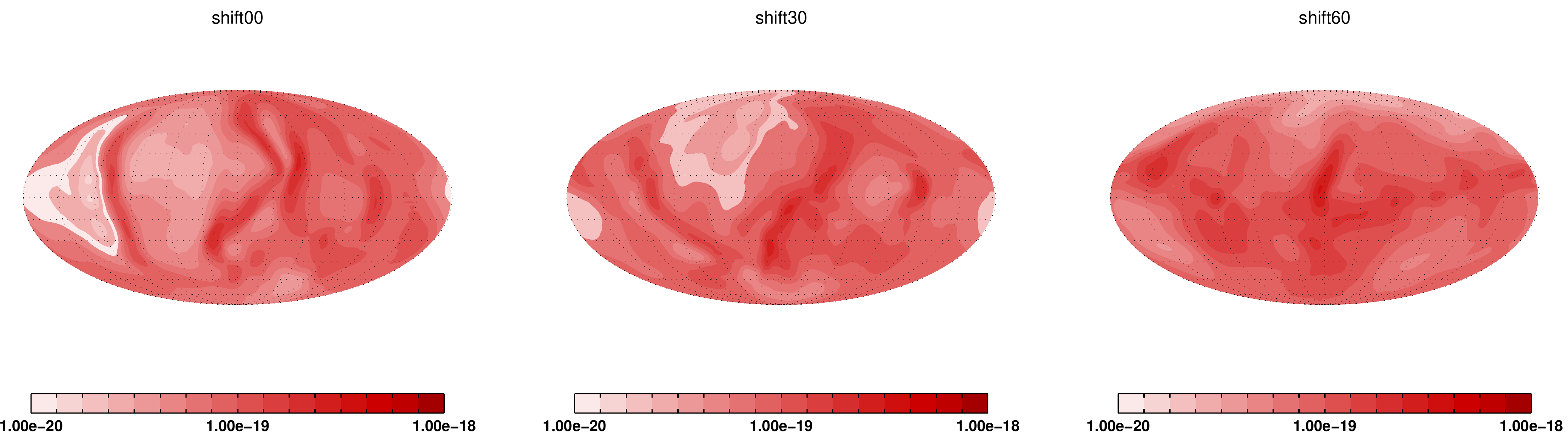}
\caption{Mass flux maps at 1 AU where we vary the sunspot latitudinal placement. All solutions are `sf2df2', for a 27 day rotation period and have ($M_{\odot}$ yr$^{-1}$ deg$^{-1}$) as unit. Note that `shift60' is the same scenario as `sf2df2' in Figure \ref{fig:massflux2}.}
\label{fig:massflux1}
\end{figure}

Before we attempt to simulate the young Sun solar wind we first ascertain how well the SC model performs for the case of the present Sun. We extract the SC solution along the trajectory of the Advanced Composition Explorer (ACE) spacecraft, which is situated at the L1 Earth-Sun Lagrange point, at $\sim\!\!1$ AU, and compare it with in-situ ACE data taken from http://cdaweb.gsfc.nasa.gov. We repeat this procedure for each simulation and obtain solar wind solutions over the course of a full solar rotation. Simulations in subsequent sections follow the same method.\\ We also test sensitivity of the solar wind solutions, i.e. of solar wind speed, density, magnetic field at 1 AU, to latitudinal placement of sunspots in the modified magnetogram. Figure \ref{fig:mgram2} shows the original and modified magnetograms used in this sensitivity analysis. We carry out several simulations with four magnetograms: `shift00', which represents the original unmodified magnetogram for CR1958; `shift30', where the sunspots have been moved to $\sim\!30^{\circ}$ latitude; `shift60', with the sunspots at $\sim\!60^{\circ}$ latitude and `CR2074-shift00', which represents a comparison magnetogram for a solar minimum. For these cases, both sunspot and dipole magnetic fields are scaled by a factor of two, which is a baseline scaling always applied to reduce discrepancies between total magnetic flux predicted by the magnetogram driven potential field model and that observed at 1 AU \citep{cohen2008}. Comparing the simulated solar wind to ACE data for the present Sun, Figure \ref{fig:scruns2} shows the predicted and observed solar wind conditions: plasma speed $u$ (km/s), magnetic field strength $B$ (nT), plasma number density $n$ (cm$^{-3}$) and the local value of the mass flux $\rho u$ ($M_{\odot}$ yr$^{-1}$ m$^{-2}$) at 1 hour averages along the location of Earth during CR1958.\\
As is clear from the ACE data the solar wind is quite variable. The record reveals two corotating interacting regions (CIR) as well as a coronal mass ejection (CME). CIRs are compression regions created when the fast solar wind exceeds the slow solar wind as the two propagate into interplanetary space. In general, two CIRs appear in a single CR, and they are associated with a smooth increase in wind speed (transition from slow to fast wind), and an increase in $B$, $n$, and $T$ due to the compression in the CIR. A CME is a solar eruption that carries with it large amount of ionized gas and is propagating through interplanetary space with a shock at its front. Therefore, a signature of a CME in the in-situ measurement will appear as a sharp increase in the plasma parameters, see the end of the record. There are some other indicators, such as composition data, for a CME in the ACE data that are not discussed, since here we are interested in the steady-state solution for the solar wind and modeling of CMEs is omitted.\\
For each observable we also show the average value for the entire year 2000. In the understanding that we are comparing a steady-state solution for the quiescent Sun to a single realization of the Sun's output we note that the solar wind solutions for the original magnetogram, `shift00', perform very well in capturing the main features with the exception of the CME which we do not model.\\
In order to safeguard against a particularly (un)fortunate comparison of the solar wind solution to the ACE data we also show how the solution varies with latitude at 1 AU. These are the `shift00-up' and `shift00-down' solutions which were taken at $\pm15^{\circ}$ away from the ecliptic plane. The solution is well behaved and does not vary significantly away from Earth's orbit. For each panel the solar wind solutions appear to slightly lead in phase. This is likely due to a slight overestimation of speed causing features to arrive at 1 AU a little sooner. For each variable the solar wind solutions also perform very well in terms of magnitude. We note that we slightly underestimate the magnetic field at 1 AU, however this is due to how we model the solar corona. Since we start from a specification of the magnetic potential field, any observed open magnetic flux originating from outside the polar coronal holes, e.g. at lower latitudes and around active regions, goes unmodelled yielding a discrepancy in magnetic field \citep[see][]{cohen2008}.\\
The mass flux panel clearly shows that density is the dominant factor with the velocity only acting as a prefactor.
As we move the sunspots to increasingly higher latitudes, `shift30' and `shift60', we see that the speed and density are quite sensitive to latitudinal sunspot placement, Figure \ref{fig:scruns2} and the magnetic field less so. For a 30 degrees shift the features in each panel are less pronounced and for a 60 degrees shift they have all but disappeared, retaining only the average magnitudes. Figure \ref{fig:massflux1} and Table \ref{tab:massflux} show how these solutions vary in terms of mass flux. We see that with increasing latitude of the sunspots the mass flux increases slightly but the distribution remains roughly unaffected.\\
For the sake of completeness we also show a solution, `cr2074-shift00', representing a solar minimum. This solution seems able to capture some of the larger scale features in the data but, as it should do, under- and overestimates the solar wind speed and density respectively.\\
What these simulations show is that the SC model of the SWMF is very capable of modeling the solar wind for the present Sun, which makes us confident that it will be able to reasonably model the young solar wind as well.\\

We also carried out simulations for the magnetogram obtained during CR2010 and obtained consistent results, but with slightly poorer agreement with observations. This was due to an uncharacteristically active Sun for that solar maximum, which featured the 2003 Halloween storms, and the associated ``contamination'' of the more steady solar wind components with CMEs that are not included in the model.
\subsection{The role of coronal base density}
\label{subsec:density}
The solar wind parameters derived from the MHD models presented here are sensitive to the assumed value of the coronal base gas density.  The density of gas in the different regions of the present-day solar corona is well-established from extreme ultraviolet spectroscopy, and ranges from $10^8$ cm$^{-3}$ in the quiet Sun and diffuse limb to typically $\sim \! 5\times 10^{9}$ cm$^{-3}$ for active regions \citep[e.g.\ see][for recent assessments]{Keenan2008,Shestov2009,Keenan2010}.  Spectra of the full-disk Sun, like spatially-unresolved stellar observations, largely reflect the properties of the brighter, more dense active regions, and \citet{Laming1995} found densities of 1--$3\times 10^9$ cm$^{-3}$ from the full-Sun EUV spectrum of \citet{Malinovsky1973}.\\
The behaviour of plasma density in solar-like stars indicates that the more active Gyr-old Sun is likely to be characterized by somewhat higher densities.  X-ray spectra from the {\it Chandra} and {\it XMM-Newton} satellites show plasma densities in solar-like activity stars, such as Procyon and $\alpha$~Cen~A and B, are similar to full-disk solar measurements at a few $10^9$ cm$^{-3}$ \citep[e.g.][]{mewe1995,schmitt1996,drake1997,ness2002}.  Towards higher activity level stars, densities transition to a few $10^{10}$~cm$^{-3}$ at base coronal temperatures of 1--$2\times 10^6$ K \citep[e.g.][]{Testa2004,Ness2004}. These measurements are based on X-ray emission from mostly closed magnetic loops, and it is not yet certain to what extent these properties and trends are shared by open field regions from which the wind largely emanates.  Nevertheless, the indication is that the coronal base density for the $\sim\! 1$ Gyr Sun would have been about a factor of ten larger than it is today. Figure \ref{fig:scruns2} shows the solar wind solution for a model in which we specify a high coronal base density of $10^{10}$ cm$^{-3}$ as compared to the otherwise used $5\times10^{8}$ cm$^{-3}$: `shift00-hd'. This solution for a zero degree shift of the sunspots shows a much smaller solar wind speed and concomitantly much higher density. As such it reveals how the solar wind responds to an increase in coronal base density and yields a possible young Sun-Earth scenario. Note the large increase in mass flux for this solution, Table \ref{tab:massflux}. \\ 

Sensitivity of the SC model to longitudinal sunspot placement, i.e. to choice of magnetogram, was addressed in \cite{cohen2008} where SC output was compared to one year of ACE data (CR1972-CR1984) during which period the Sun was experiencing solar maximum conditions. The model was shown to perform well and captured the salient features of the solar wind such as the periodic corotating interacting regions and transients from slow to fast solar wind. As such we are confident our choice of magnetogram does not bias our solar wind solutions.
\subsection{Sunspot and dipole magnetic field}
\begin{figure}
\noindent\includegraphics[width=39pc]{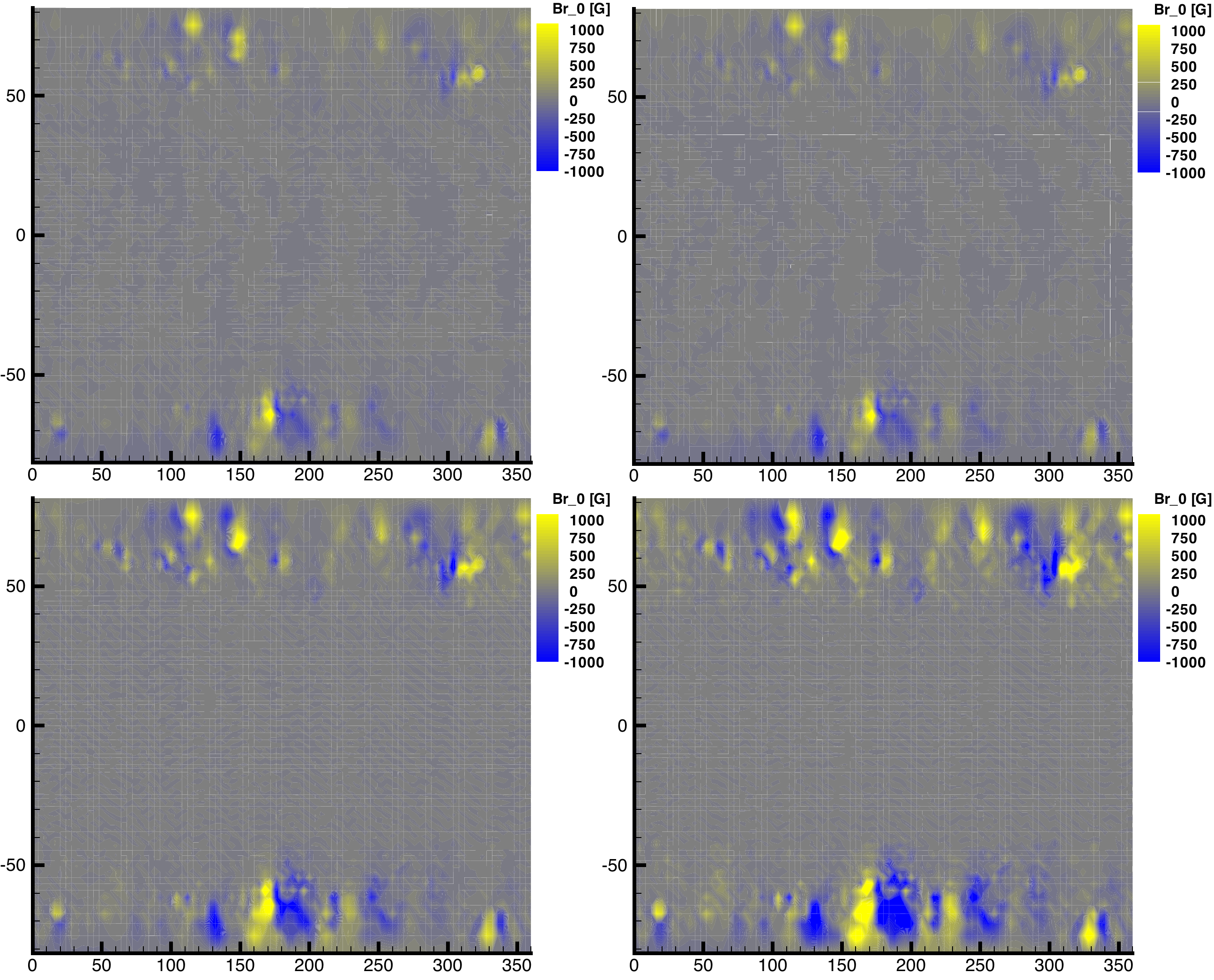}
\caption{Magnetograms used in Figure \ref{fig:scruns3}. Modified MDI magnetograms (CR1958) used as input to the Solar Corona module of SWMF to find a solution for the solar wind. We vary the sunspot magnetic field strength (sf) and the background dipole magnetic field strength (df) by a scaling factor. All magnetograms have sunspots placed at $\pm 60^{\circ}$ latitude and are for a 27 day rotation. Top-left: SF2DF5; top-right: SF2DF10; bottom-left: SF5DF2; bottom-right: SF10DF2.}
\label{fig:mgram3}
\end{figure}

\begin{figure}
\noindent\includegraphics[width=39pc]{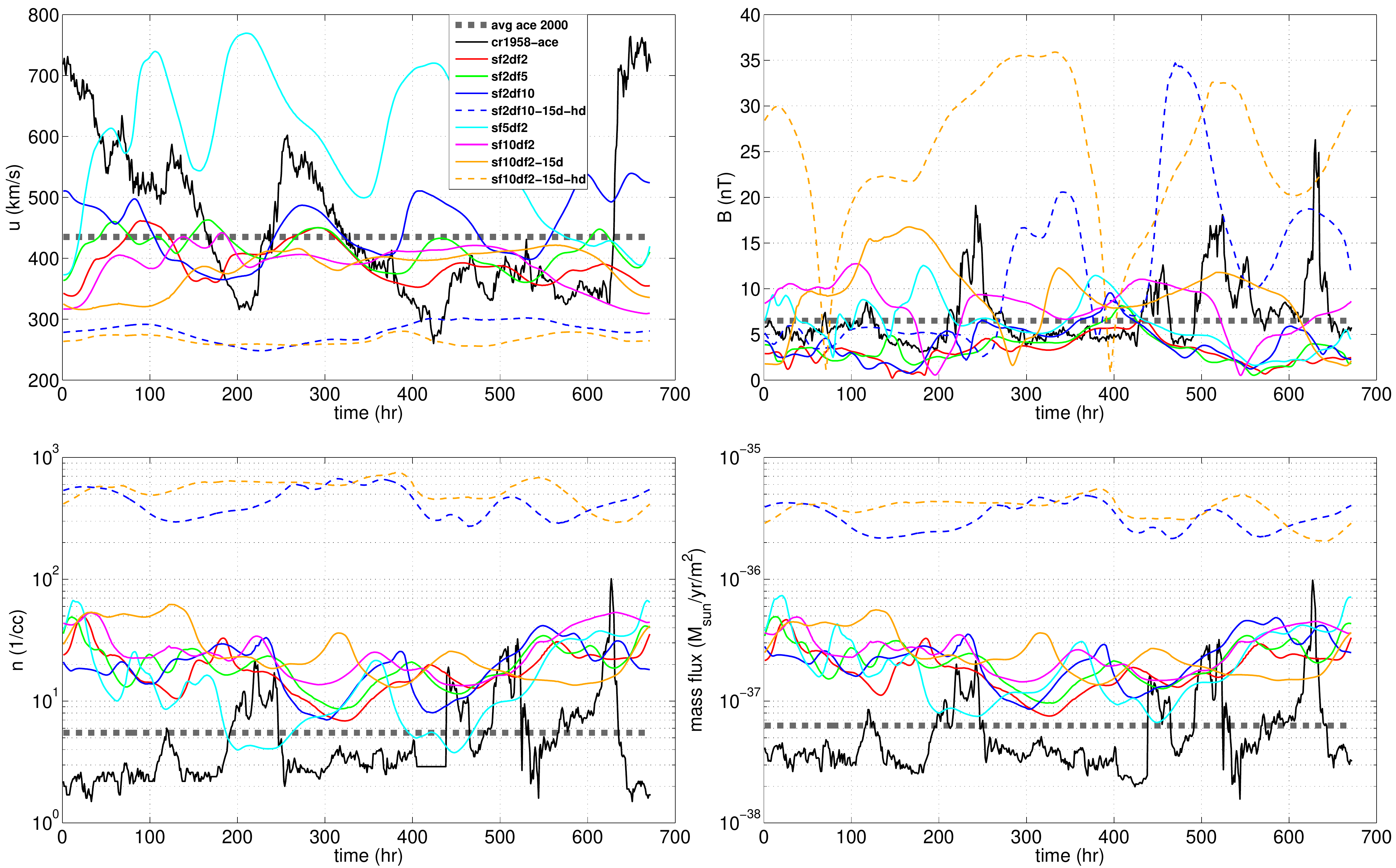}
\caption{Eight SC solar wind solutions at 1 AU where we vary the sunspot magnetic field scaling factor (sf) and the dipole magnetic field scaling factor (df). All solutions are for a magnetogram with sunspots placed at 60 degrees latitude and for a 27 day rotation period unless otherwise indicated by `15d' meaning a 15 day rotation. As a reference we show ACE data at 1 AU for CR1958 as well as the average values of ACE data for 2000.}
\label{fig:scruns3}
\end{figure}

\begin{figure}
\noindent\includegraphics[width=39pc]{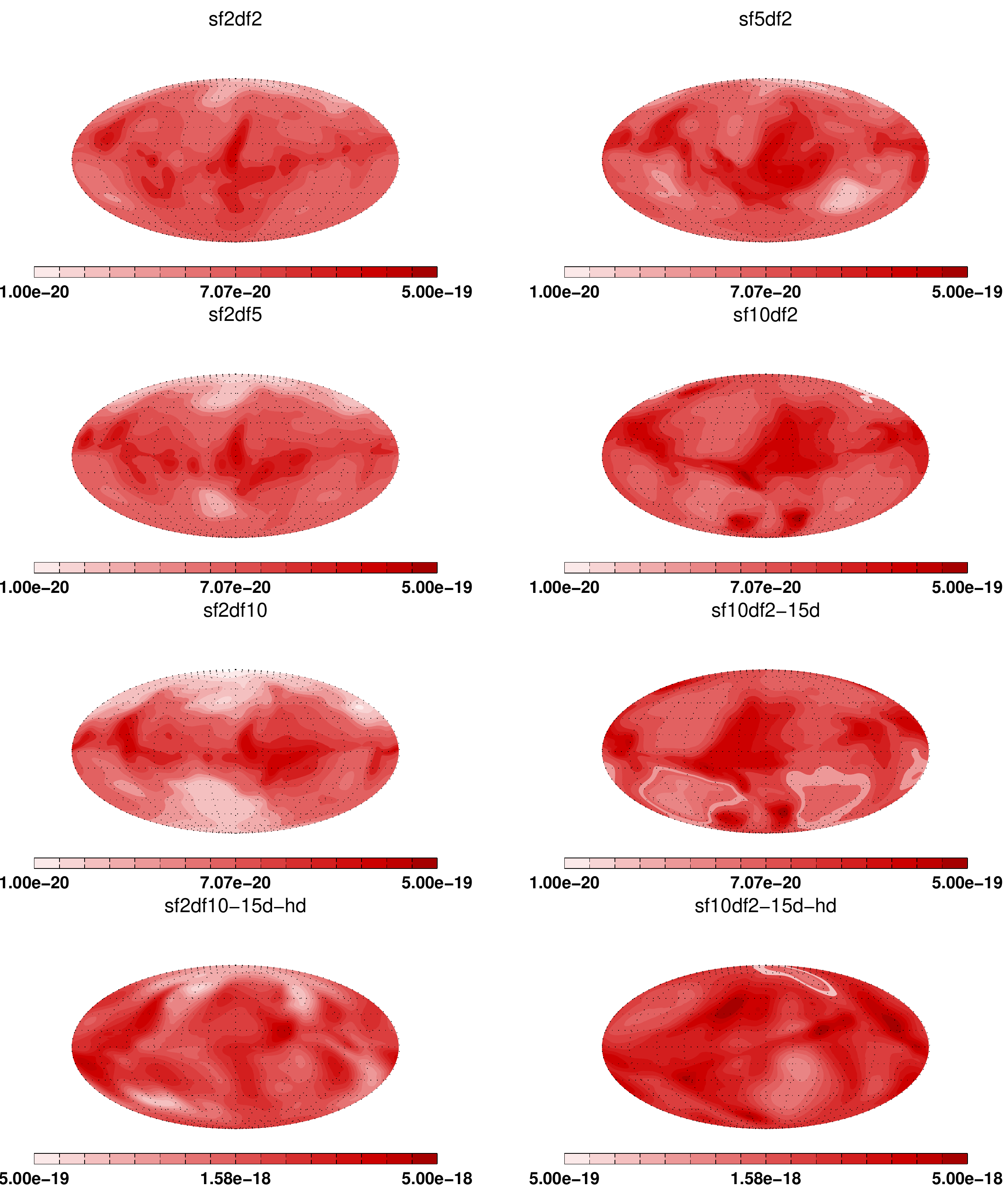}
\caption{Mass flux maps at 1 AU where we vary the sunspot magnetic field scaling factor `sf', right column, and the dipole magnetic field scaling factor `df', left column. All solutions except for the last are for a 27 day rotation period unless otherwise indicated by `15d' meaning a 15 day rotation, and have ($M_{\odot}$ yr$^{-1}$ deg$^{-1}$) as unit. Note that the last row has different limits. }
\label{fig:massflux2}
\end{figure}

Let us now look at solar wind solutions for which we independently vary the sunspot magnetic field, `sf', and the background dipole magnetic field, `df'. 
 In general, the dipole component of the solar magnetic field generates fast wind originating from regions of open field lines at high latitudes (coronal holes), and slow wind originating from open field lines at the boundary of the large helmet streamers. On top of this ambient, large-scale, bi-modal wind structure, smaller scale modulations appear due to solar wind that comes from open field lines at the vicinity of the spots. The mass flux from these regions is usually higher than the mass flux associated with the high-latitude fast wind since the source regions near the spots are denser than the ambient coronal holes. In the case of surface field dominated by the dipole component, the mass flux distribution is dominated by the bi-modal ambient wind structure so that the mass flux distribution is bi-modal as well. In the case of strong spots however, local variations in the mass flux distribution due to the spots are bigger. \\
Figure \ref{fig:mgram3} shows the magnetograms that were used as inputs to the SC model. For each magnetogram we placed the sunspots at $60^{\circ}$ latitude and scaled the sunspot and dipole magnetic fields. We use the nomenclature `sfxdfy' where `x' and `y' are scaling factors.  In Figure \ref{fig:scruns3} we show solar wind solutions for these magnetograms. All solutions are for a rotation period of 27 days, with the exception of those that have `15d' in their name, which are for 15 days. Again we plot $u$, $n$, $B$ and mass flux and show ACE data for reference. For the first three solutions, `sf2df2',`sf2df5' and `sf2df10', we only increase the dipole magnetic field which, on average, results in an increase in speed and a slight increase in plasma density. These scaling factors were guided by the \cite{petit2008} magnetic field maps for solar analogs mentioned above, that indicate a higher poloidal field by an order of magnitude than the current Sun for a shorter rotation period of 12 days that is more appropriate to the young Sun modeled here.
The magnetic field strength stays approximately the same as does the local mass flux. Looking at the mass fluxes for these runs, see Figure \ref{fig:massflux2} (left column) and Table \ref{tab:massflux}, we note that the mass flux slightly decreases with increasing dipole strength and has a tendency to decrease near high latitudes and concentrate at lower latitudes at 1 AU. \\

Next we vary only the sunspots, increasing the spot magnetic field by factors of 2, 5 and 10 (models `sf2df2', `sf5df2' and `sf10df2'; see Figure \ref{fig:mgram3}). The contours in this figure are saturated to clarify the increase in magnetic field. The magnetic field strength for `sf5df2' and `sf10df2' reaches several kG, in accordance with observations of stellar magnetic field for young solar analogs \citep{gudel2007}. As Figure \ref{fig:massflux2} (right column) and Table \ref{tab:massflux} indicate, at 1 AU this increases the solar wind speed more severely,  the magnetic field intensifies, the density slightly decreases and the local mass flux sees slightly stronger variations. 

The mass flux is enhanced by increasing the sunspot magnetic field and becomes concentrated at larger local regions comparing to the strong dipole cases. The wind speed increases for `sf5df2' but decreases for `sf10df2'. This is due to the fact that when the high-latitude spots become strong enough, they close down high-latitude open field lines. This process eliminates much of the fast wind originated from coronal holes and associated with lower mass flux, and fill the interplanetary space with wind associated with the vicinity of the active regions.\\

As noted earlier, because solar-like stars lose angular momentum over time to their winds, they will have had shorter rotation periods in the past. We tested the impact of this by repeating the last solar wind model with a 15 day rotation period, `sf10df2-15d', the main result of which is that the solution is slightly phase shifted with respect to the same solution for 27 days, `sf10df2'.\\ 
The solar wind solutions `sf2df10$\_$15d$\_$hd' and `sf10df2$\_$15d$\_$hd' feature a strong dipole or sunspot magnetic field respectively, a short rotation period and a high coronal base density in an effort to determine their combined effects on the solar wind and obtain a realistic possible scenario for the young Sun. In keeping with the solution `shift00-hd' of the previous section, Figure \ref{fig:scruns3} and Table \ref{tab:massflux} show a relatively slow solar wind, high wind density, a stronger magnetic field and naturally a high (local) mass flux compared to similar non-hd solutions. Figure \ref{fig:massflux2} also shows this on the bottom row where we point out the higher mass flux limits on the colorbar. Although the mass flux distribution for the `hd' solutions does not reveal any noteworthy specific features, it is clear that the coronal base density is a dominant term controlling the solar wind mass flux.\\
We note that all solutions shows a strong anti-correlation between speed and density as expected \citep{mccomas2007}.\\

In summary, we have parameterized the SC model to obtain a reasonable agreement with present 1 AU data.  Observations of young stars reveal that stellar spots are located at higher latitudes \citep{hussain2002} and so we have shifted the spots by 30$^{\circ}$ and 60$^{\circ}$ to determine the effect on the mass flux distribution. We also used the SC model to test how variations of the magnitude of the sunspots' magnetic field, as well as the ambient weak field affect the mass flux distribution. In so doing we have covered a comprehensive set of scenarios as far as the surface magnetic field distribution is concerned. This also gives us constraints on the mass flux, which is based on a detailed model parameterization instead of on scaling relationships such as those by \cite{wood2002, wood2005}. As mentioned before, \cite{zieger2004, vogt2004, zieger2006} carried out comprehensive parameter studies in which they mainly varied dipole moment magnitude and tilt but, with the exception of the IMF, kept the solar wind parameters constant. Here we have obtained a range of solar wind scenarios which we can now use to force Earth's paleomagnetosphere with. 

\section{Other mass flux indications}
\label{sec:massflux}
\cite{wood2002, wood2005} used Ly$\alpha$ spectra to infer hot hydrogen absorption in astrospheres - the stellar analogs of the heliosphere - which they used diagnostically to determine stellar mass loss rates. To quantitatively measure mass-loss rates based on astrospheric absorption requires the use of hydrodynamic models of the astrosphere, which in turn are based on a model of the Sun's heliosphere that reproduces heliospheric HI absorption. This requires assumptions regarding: solar wind velocity, proton density and temperature, in addition to local interstellar medium (ISM) properties. Application to other stars relies on ISM properties not varying significantly between the different stars studied. Furthermore, an additional, and perhaps dominant source of error in HI absorption measurements comes from the need to assume a form for the unabsorbed stellar Ly$\alpha$ profile. \cite{wood2002,wood2005} estimated a relationship between the stellar mass loss rate $\dot{M}$ and X-ray flux $F_X$
 \begin{equation}
\label{eq:xray}
\dot{M}\propto F_X^{1.34\pm 0.18}
\end{equation} 
This can be combined with estimates from \cite{ayres1997} of rotation rates, $V_{rot}$, for stellar analogs \citep{wood2005}
 \begin{equation}
\label{eq:vrot}
V_{\mathrm{rot}}\propto t^{-0.6\pm0.1}
\end{equation} 
which shows that as stars age their rotation rates become smaller, due to magnetic braking, and
 \begin{equation}
\label{eq:fx}
F_X\propto V_{\mathrm{rot}}^{2.9\pm0.3}
\end{equation} 
which illustrates that a smaller rotation rate means less magnetic activity and concomitantly less X-ray flux. The combination of equations \ref{eq:xray}, \ref{eq:vrot} and \ref{eq:fx} yields
 \begin{equation}
\dot{M}\propto t^{-2.33\pm0.55}
\label{eq:massloss}
\end{equation} 
Using equations \ref{eq:xray} and \ref{eq:massloss} \cite{tarduno2010} arrived at a range of mass loss rates between 2.4$\times10^{-13}$ to 1.5$\times10^{-12}$ $M_{\odot}$ yr$^{-1}$ for the Archean Earth. \\ \cite{holzwarth2007} used a different approach, based on a magnetized solar wind model assuming power laws for the dependence of thermal and magnetic wind parameters on the stellar rotation rate,  to arrive at a mass flux for cool main sequence stars of no more than 2$\times 10^{-13}$ $M_{\odot}$ yr$^{-1}$, i.e. less than 10 times the present solar mass loss rate of 2.1$\times 10^{-14}$ $M_{\odot}$ yr$^{-1}$ \citep{feldman1977,gudel2007}. \\
With the exception of the solar wind solutions with high coronal base densities, `shift00$\_$hd', `sf2df10$\_$15d$\_$hd' and `sf10df2$\_$15d$\_$hd', which have mass fluxes between $\sim\! 1-1.5\times10^{-12}$ $M_{\odot}$ yr$^{-1}$, our solutions have mass fluxes within the range of 5 - 8$\times10^{-14}$ $M_{\odot}$ yr$^{-1}$, well below the literature estimates by an order of magnitude. From this we may infer that either scaling relationships such as those derived by \cite{wood2002,wood2005} and \cite{holzwarth2007} are overestimating young solar mass flux, or that higher coronal base densities are warranted in our models of the young Sun's solar wind. Given the expectation that younger solar-like stars are characterized by higher coronal base densities, see subsection \ref{subsec:density}, the latter seems more likely. However, we also note that the astrospheric response to stellar winds on which the \cite{wood2002,wood2005} mass loss estimates are based has a timescales of years or more.  As such, these estimates include the average effect of mass loss through CMEs. The partition of mass loss between the steady wind and more impulsive CMEs is not known for stars significantly more active than the Sun, but based on the higher flaring rate of young solar analogs it is reasonable to expect that a greater proportion of the mass loss could occur impulsively. It is plausible that the difference in mass loss rates for our steady wind models and the estimates of \cite{wood2002,wood2005} is due to the impulsive component. 

\section{Forcing Earth's paleomagnetosphere}
\label{sec:earth}
In order to assess the effect of various solar wind scenarios on the Earth's paleomagnetosphere we rely on the Global Magnetosphere (GM) MHD model of the SWMF, the input parameters of which we modify to simulate a paleomagnetosphere. Based on the solar wind solutions of section \ref{sec:sun}, using the time-averaged values of Table \ref{tab:massflux}, we develop several scenarios with which to simulate Earth's paleomagnetospheric response to the young Sun's possible forcing (see Table \ref{tab:gmcases}). The GM domain ranges from about 30 $R_E$ on the dayside to about 220 $R_E$ on the nightside and about 130 $R_E$ perpendicularly to the Sun-Earth line. The inner boundary of the GM is 2.5 $R_E$ from the center of Earth with its boundary conditions set by the Ionospheric Electrodynamics (IE) model of the SWMF, \citep{ridley2002,ridley2005}.  We use three levels of grid refinement in the GM model with a minimum grid size of 1/8 $R_E$ closest to the Earth and a maximum grid size of 8 $R_E$.\\ 
 
Although the Earth's magnetic field is not strictly dipolar, over geologic timescales it does average to a dipole field with its axis of symmetry coincident with Earth's rotation axis \citep{merrill1983}. As the magnetic field reverses polarity its morphology changes as does its dipole axis which can experience excursions exceeding 45$^{\circ}$. In this study we assume a dipolar magnetic field and a number of dipole tilt scenarios. \\ 
\cite{tarduno2010} have made paleomagnetic measurements of 3.5 Ga old samples that indicate a virtual dipole moment (VDM) $\sim$50-70$\%$ of the present dipole moment, but representing field averages spanning only decades to centuries. We must be cautious when using VDM's as representative of actual dipole moments, especially when the measurements the VDM  is based on all come from the same, or similarly placed, sites, in which case the spatial distribution of the measurements is not adequate to separate the dipole from non-dipolar components. As such, VDM's may under- or overestimate the actual dipole moment.\\ Given the geomagnetic reversal rate over the last 160 Ma of $\sim$5 per million years, with an average reversal duration of $10^3 - 10^4$ years \citep{mcfadden2000} and assuming this reversal rate has not changed significantly over the last 3.5 Ga we estimate that the (purely statistical) probability of the aforementioned measurements  of sampling normal, non-transient, field conditions is between 95-99.5$\%$.  However dynamo theory suggests a scaling relationship that indicates that magnetic field strength inversely scales with the square root of the rotation period, implying a stronger paleomagnetic field. \\ 
Planetary dynamos are believed to exist in a regime where the Coriolis and Lorentz forces are balanced by each other, which is captured by the Elsasser number
 \begin{equation}
\Lambda=\frac{\sigma B^2}{\rho \Omega},
\end{equation} 
a ratio of the two, and should always be of order unity \citep{stevenson2003}. Here $\sigma$ is the electrical conductivity and $\rho$ the density. We can obtain a scaling relationship between the magnetic field strength $B$ and rotation period $T=2\pi/\Omega$, where $\Omega$ is the rotation rate.
 \begin{equation}
B\propto T^{-1/2}
\end{equation} 
\cite{hansen1982} suggests a rotation period of $\sim\!\!15$ hours for the Archean Earth, due to the closer proximity of the Moon, which implies a $\sim \!\!30\%$ increase in the magnetic field strength. Here we account for both possible scenarios: a stronger dipole field that scales with the rotation period and a weaker field as indicated by paleomagnetic measurements.\\

We assess each GM solution in terms of several observables. The subsolar magnetopause stand-off distance, $R_{mp}$, is determined by the position of minimum plasma speed at the stagnation point on the Sun-Earth line. The magnetopause dawn-dusk flank distance, $R_{dd}$, and the north-south flank distance, $R_{ns}$, are determined by finding the maximum current density on the magnetopause. We assess the extent of the polar cap by inspecting the boundary between open and closed field lines at the inner boundary of the simulation, the IE grid. We find the polar cap magnetic co-latitude, $\theta_{pc}$, for the dayside and nightside. We also calculate the polar cap area relative to that of a GM solution for the present Earth. \\

In scenarios A-C in Table \ref{tab:gmcases} we vary Earth's dipole magnetic field from 25-133$\%$ of the present field. Solar wind speed and density are based on the young Sun solutions with strong sunspot magnetic fields, `sf10df2' and `sf10df2-15d'. Figure \ref{fig:mspheres}, top-row, shows how the paleomagnetosphere changes with decreasing terrestrial dipole field strength. Qualitatively, it is clear that the magnetopause stand-off distance strongly decreases and the magnetosphere becomes very compressed. Quantitatively, Table \ref{tab:gmcases} shows that $R_{mp}$ more than halves, with strongly decreasing $R_{dd}$ and $R_{ns}$ and increasing polar cap.\\

During a geomagnetic polarity reversal the dipole field can exhibit several transitions, \citep{zieger2004,gubbins1994}. The field may become strongly multipolar before it settles into a dipole field of the opposite polarity, the dipole field may severely weaken and gain in strength with the opposite polarity or the dipole field may experience several excursions, where the dipole axis tilts, before it finally flips. We address this latter possibility in scenarios D-F, which are based on the first four solutions of Table \ref{tab:massflux}, `sf2df2' - `sf5df2'. The middle-row of Figure \ref{fig:mspheres} shows the response to these scenarios and from Table \ref{tab:gmcases} it is clear that while $R_{mp}$, $R_{dd}$ and $R_{ns}$ do not change much with dipole tilt, although $R_{ns}$ does become asymmetric, the polar cap becomes much smaller.  \\

In scenario G we run a similar scenario to D-F but for a magnetic field similar to that measured in \cite{tarduno2010}. We obtain a $R_{mp}$ of 6 $R_E$ which is just within the upper margin of errror of their estimate. \\

Earth's magnetosphere is not constantly exposed to a southward IMF along with the subsequent magnetic reconnection and magnetic field erosion. The effect of a northward IMF is shown with scenario H where we see that the magnetopause stand-off distance is relatively greater, 6.25 $R_E$, than that for the previous scenario G, but the main effect is that there is no polar cap of any significance, i.e. the absence of reconnection prohibits the solar wind from directly accessing the upper atmosphere. We have also run a northward IMF scenario for a strong dipole field, not shown here, which yielded $R_{mp}\sim\!9$ $R_E$. \\

Scenarios I through M, based on solar wind solutions `shift00-hd' (I), `sf2df10-15d-hd' (J, K) and `sf10df2-15d-hd' (L, M), investigate the effects of a higher coronal base density, which is characterized by a slow solar wind, a high wind density and a large mass flux (compared with present day). We also vary Earth's magnetic field between 50 and 133 $\%$ of the present field. The effects of these scenarios on the magnetosphere are stand-off distances between 4 and 6 $R_E$ depending on Earth's magnetic field strength, flank distances between 5 and 9 $R_E$, very high plasma pressures and relatively larger polar caps, see Figure \ref{fig:mspheres} which illustrates this with scenarios J, K and L, as well as Table \ref{tab:gmcases}. However, this $\sim\!15-50\%$ increase in polar cap is much less compared with polar cap estimates of \cite{tarduno2010}. They used a scaling relationship from \cite{siscoe1975}
 \begin{equation}
\cos(\theta_{pc})=\left(\frac{M}{M'} \right)^{-1/6}P^{1/12}\cos(\theta_{pc}')
\label{eq:pc}
\end{equation} 
where $\theta_{pc}'$ is the present latitude of the division, $P$ is the solar wind dynamic pressure normalized at the magnetopause to its present value of $\sim\!\!2$ nPa and $M$, $M'$ are the Earth's dipole moments in the Archean and the present respectively. \cite{tarduno2010} suggested a polar cap increase of $\sim\!300\%$ with respect to present conditions. At present this discrepancy is unresolved. However it might be related to the inherent assumption of a circular polar cap in equation \ref{eq:pc}. Indeed Table \ref{tab:gmcases} does show different values of $\theta_{pc}$ for the day- and nightside indicating non-circular polar caps. \\

Finally in scenario N we show the reference scenario of the present Sun-Earth interaction.

\section{Discussion and conclusion}
\label{sec:conclusion}
We have carried out a study of the young Sun's solar wind and its interaction with Earth's paleomagnetosphere. Using numerical MHD simulations we have modeled the solar wind and evaluated its sensitivity to sunspot placement, strength of sunspots' magnetic field, strength of dipole field, coronal base density and rotation period. We have parameterized our solar wind model so as to agree with present solar wind data taken at 1 AU. We have also analyzed how sensitive the solar mass flux distribution is to the aforementioned parameters. In comparing these findings to scaling relationships found in the literature we have determined that our mass flux estimates for the young Sun are approximately an order of magnitude less, with the exception of those solar wind solutions that have a higher coronal base density as their initial condition. We have found that the coronal base density is the controlling factor in mass flux magnitude, whereas mass flux distribution seems mainly affected by sunspot and background dipole magnetic field strength. Sunspot latitudinal placement seems to affect mass flux slightly but leaves mass flux distribution unaffected. \\
Given literature estimates as well as observations we conclude that simulations of the young Sun's solar wind likely require a high coronal base density.\\

Based on our solar wind simulations we have established a number of solar wind scenarios with which to force the Earth's paleomagnetosphere. In these scenarios we varied parameters such as the Earth's dipole field strength and tilt, the solar wind speed, density, IMF orientation and strength. We have found that for a solar wind with high density, strong southward IMF and very weak dipole field the magnetopause stand-off distance can be as small as 4.25 $R_E$. For a strong dipole field this increases to 8.5 $R_E$ and up to 9 $R_E$ for a northward IMF. 

\begin{figure}
\noindent\includegraphics[width=30pc]{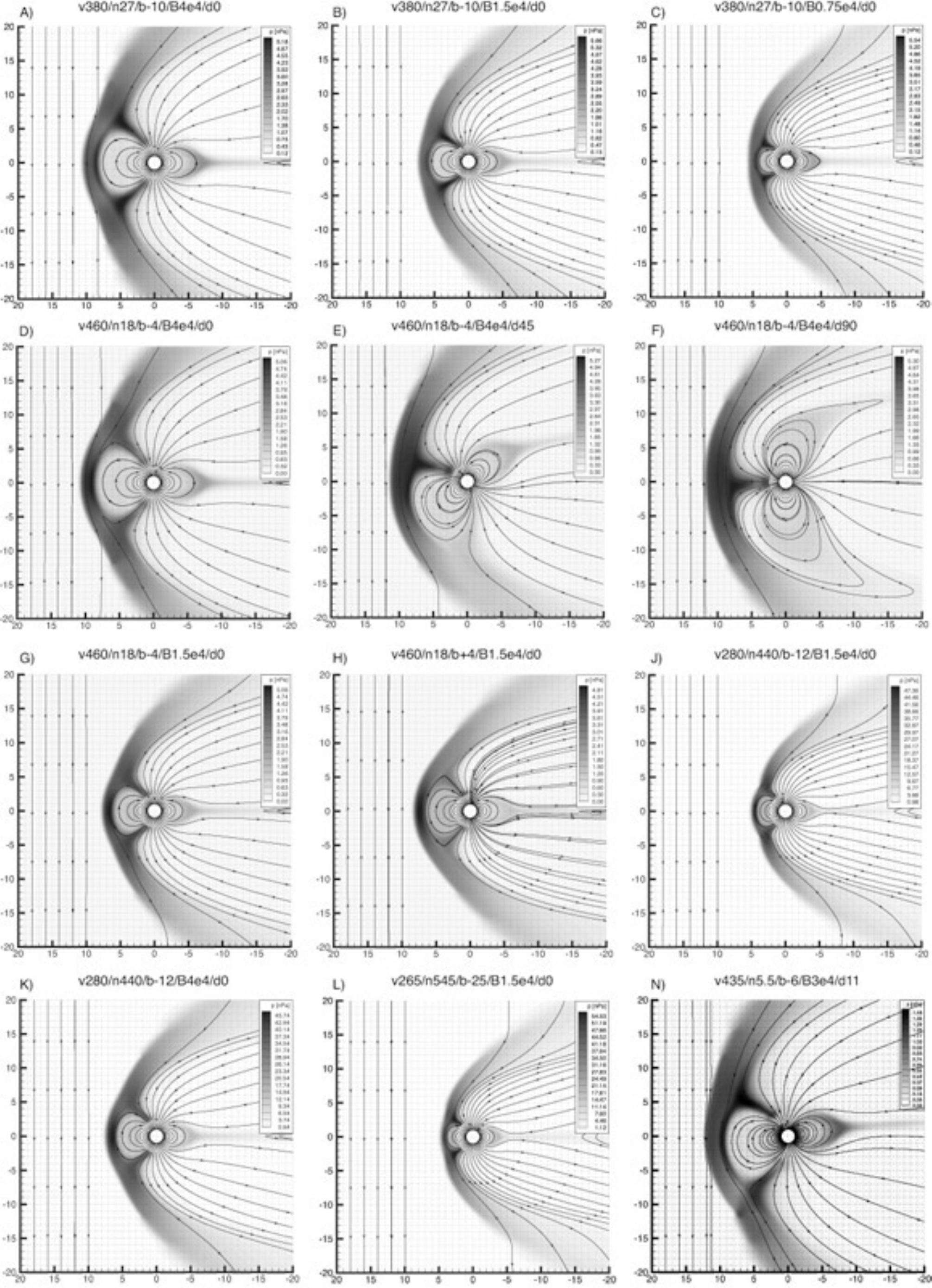}
\caption{GM solutions for various parameter choices. Shown are cross-sections of the magnetosphere along the noon-midnight plane. The relevant parameters are indicated above each figure: `plasma speed/plasma number density/IMF z-component/Earth's dipole magnetic field/dipole tilt. The top row varies Earth's magnetic field from 133$\%$ to 25$\%$ of the present dipole. The second row varies dipole tilt from $0^{\circ}$ to $90^{\circ}$ (`pole-on'). The third row shows from left to right a solution based on `sf2df2' - `sf5df2' with a weak dipole field, a solution with northward $b^{imf}_z$, and a solution based on the `sf2df10$\_$15d$\_$hd' solar wind solution. The bottom row shows respectively solutions based on: `sf2df10$\_$15d$\_$hd' with a stronger Earth magnetic field, `sf10df2$\_$15d$\_$hd' and a solution representing present Sun-Earth interaction. }
\label{fig:mspheres}
\end{figure}

It is clear from our simulations that the young Sun's solar wind will have had easier access to Earth's paleoatmosphere, especially in the case of a relatively weaker, with respect to the present, magnetic field: compare for instance scenarios L and N in Figure \ref{fig:mspheres}. As such, assuming a weak terrestrial magnetic field, our findings support those made by \cite{tarduno2010}, with the exception of their polar cap calculation. Using an uncomplicated power-law based approach they found a greatly compressed paleomagnetosphere as well yielding obvious implications for heating and subsequent expansion of Earth's exosphere with important ramifications for Earth's early atmospheric evolution.\\

We note however that convergence of the stellar mass loss rate scaling relationships by  \cite{wood2002, wood2005}, which \cite{tarduno2010} based some of their conclusions on, with our solar wind simulations for high coronal base densities is not conclusive evidence that either are correct. Measurements of stellar coronal base densities, for young solar analogs, are sparse and the myriad uncertainties that accompany the aforementioned scaling relationships are well documented. Until more and better observations are made of stellar mass loss rates, stellar coronal base densities and/or the number of assumptions and uncertainties are reduced in the scaling relationships, our simulations of the paleomagnetospheric response to the young Sun's solar wind give the most realistic results to date.\\

We emphasize that this study was carried out for the quiescent Sun and modeled only steady-state interactions. As such, transient, dynamic interactions between the Sun and Earth's paleomagnetosphere have not been modeled. As noted by \cite{tarduno2010}, the magnetosphere could be more compressed during strong CME events, that are expected to be much more frequent and energetic on a young solar-like star, than our steady-state scenarios suggest. In this context, it will be important to understand the relative contributions of steady solar winds and impulsive events to the mass loss experienced by more active solar analogs. On the other hand, assuming a close correlation between the CME source region and the distribution of active regions on the solar disk  \citep{gopalswamy2008}, we cannot dismiss the possibility that most CMEs at that time came from high latitudes on the disk and therefore the number of CMEs actually reaching the Earth was in fact smaller than at the present time.\\

Having obtained a range of magnetosphere sizes based on physically realistic solar wind scenarios, this study provides a well-grounded framework within which to undertake further paleomagnetospheric and paleoatmospheric evolution studies.


%
%
%
%
%
%

%
%
%
%

\begin{acknowledgments}
We are grateful to Dimitar Sasselov, Itay Halevy and Alex Glocer for insightful discussions. OC is supported by SHINE through NSF ATM-0823592 grant. JJD was funded by NASA contract NAS8-39073 to the \emph{Chandra X-ray Center}. Simulation results were obtained using the Space 
Weather Modelling Framework, developed by the Center for Space Environment Modelling, at the University of Michigan with funding support from NASA ESS, NASA ESTO-CT, NSF KDI, and DoD MURI.  The GM computations in this paper were run on the Odyssey cluster supported by the Harvard FAS Sciences Division Research Computing Group.

\end{acknowledgments}

%
%

\begin{thebibliography}{56}
\expandafter\ifx\csname natexlab\endcsname\relax\def\natexlab#1{#1}\fi
\expandafter\ifx\csname url\endcsname\relax
  \def\url#1{\texttt{#1}}\fi
\expandafter\ifx\csname urlprefix\endcsname\relax\def\urlprefix{URL }\fi

\bibitem[{{Arge} and {Pizzo}(2000)}]{arge2000}
{Arge}, C.~N., {Pizzo}, V.~J., May 2000. {Improvement in the prediction of
  solar wind conditions using near-real time solar magnetic field updates}. J.
  Geophys. Res. 105, 10465--10480.

\bibitem[{{Ayres}(1997)}]{ayres1997}
{Ayres}, T.~R., Jan. 1997. {Evolution of the solar ionizing flux}. J. Geophys.
  Res. 102, 1641--1652.

\bibitem[{{Cohen} et~al.(2007){Cohen}, {Sokolov}, {Roussev}, {Arge},
  {Manchester}, {Gombosi}, {Frazin}, {Park}, {Butala}, {Kamalabadi}, and
  {Velli}}]{cohen2007}
{Cohen}, O., {Sokolov}, I.~V., {Roussev}, I.~I., {Arge}, C.~N., {Manchester},
  W.~B., {Gombosi}, T.~I., {Frazin}, R.~A., {Park}, H., {Butala}, M.~D.,
  {Kamalabadi}, F., {Velli}, M., Jan. 2007. {A Semiempirical
  Magnetohydrodynamical Model of the Solar Wind}. Astrophysical Journal,
  Letters. 654, L163--L166.

\bibitem[{{Cohen} et~al.(2008){Cohen}, {Sokolov}, {Roussev}, and
  {Gombosi}}]{cohen2008}
{Cohen}, O., {Sokolov}, I.~V., {Roussev}, I.~I., {Gombosi}, T.~I., Mar. 2008.
  {Validation of a synoptic solar wind model}. J. Geophys. Res. 113~(A12),
  3104--+.

\bibitem[{{Drake} et~al.(1997){Drake}, {Laming}, and {Widing}}]{drake1997}
{Drake}, J.~J., {Laming}, J.~M., {Widing}, K.~G., Mar. 1997. {Stellar Coronal
  Abundances. V. Evidence for the First Ionization Potential Effect in alpha
  Centauri}. \apj 478, 403--+.

\bibitem[{{Durney} and {Latour}(1978)}]{durney1978}
{Durney}, B.~R., {Latour}, J., 1978. {On the angular momentum loss of late-type
  stars}. Geophysical and Astrophysical Fluid Dynamics 9, 241--255.

\bibitem[{{Feldman} et~al.(1977){Feldman}, {Asbridge}, {Bame}, and
  {Gosling}}]{feldman1977}
{Feldman}, W.~C., {Asbridge}, J.~R., {Bame}, S.~J., {Gosling}, J.~T., 1977.
  {Plasma and Magnetic Fields from the Sun}. In: {O.~R.~White} (Ed.), The Solar
  Output and its Variation. pp. 351--+.

\bibitem[{Glatzmaier and Roberts(1995)}]{glatzmaier1995}
Glatzmaier, G., Roberts, P., 1995. A three-dimensional self-consistent computer
  simulation of a geomagnetic field reversal. Nature 377, 203--209.

\bibitem[{{Gopalswamy} et~al.(2008){Gopalswamy}, {Akiyama}, {Yashiro},
  {Michalek}, and {Lepping}}]{gopalswamy2008}
{Gopalswamy}, N., {Akiyama}, S., {Yashiro}, S., {Michalek}, G., {Lepping},
  R.~P., Feb. 2008. {Solar sources and geospace consequences of interplanetary
  magnetic clouds observed during solar cycle 23}. Journal of Atmospheric and
  Solar-Terrestrial Physics 70, 245--253.

\bibitem[{{Grie{\ss}meier} et~al.(2004){Grie{\ss}meier}, {Stadelmann}, {Penz},
  {Lammer}, {Selsis}, {Ribas}, {Guinan}, {Motschmann}, {Biernat}, and
  {Weiss}}]{griessmeier2004}
{Grie{\ss}meier}, J., {Stadelmann}, A., {Penz}, T., {Lammer}, H., {Selsis}, F.,
  {Ribas}, I., {Guinan}, E.~F., {Motschmann}, U., {Biernat}, H.~K., {Weiss},
  W.~W., Oct. 2004. {The effect of tidal locking on the magnetospheric and
  atmospheric evolution of ``Hot Jupiters''}. \aap 425, 753--762.

\bibitem[{Gubbins(1994)}]{gubbins1994}
Gubbins, D., 1994. Geomagnetic polarity reversals: A connection with secular
  variation and core?mantle interaction? Reviews of Geophysics 32(1), 61--83.

\bibitem[{{G{\"u}del}(2007)}]{gudel2007}
{G{\"u}del}, M., Dec. 2007. {The Sun in Time: Activity and Environment}. Living
  Reviews in Solar Physics 4, 3--+.

\bibitem[{{Guinan} and {Engle}(2009)}]{guinan2009}
{Guinan}, E.~F., {Engle}, S.~G., Jun. 2009. {The Sun in time: age, rotation,
  and magnetic activity of the Sun and solar-type stars and effects on hosted
  planets}. In: {E.~E.~Mamajek, D.~R.~Soderblom, \& R.~F.~G.~Wyse} (Ed.), IAU
  Symposium. Vol. 258 of IAU Symposium. pp. 395--408.

\bibitem[{{Guzik} et~al.(1987){Guzik}, {Willson}, and {Brunish}}]{guzik1987}
{Guzik}, J.~A., {Willson}, L.~A., {Brunish}, W.~M., Aug. 1987. {A comparison
  between mass-losing and standard solar models}. Astrophysical Journal. 319,
  957--965.

\bibitem[{{Hansen}(1982)}]{hansen1982}
{Hansen}, K.~S., Aug. 1982. {Secular effects of oceanic tidal dissipation of
  the moon's orbit and the earth's rotation}. Reviews of Geophysics and Space
  Physics 20, 457--480.

\bibitem[{{Holzwarth} and {Jardine}(2007)}]{holzwarth2007}
{Holzwarth}, V., {Jardine}, M., Feb. 2007. {Theoretical mass loss rates of cool
  main-sequence stars}. Astron. and Astrophys. 463, 11--21.

\bibitem[{{Hussain} et~al.(2002){Hussain}, {van Ballegooijen}, {Jardine}, and
  {Collier Cameron}}]{hussain2002}
{Hussain}, G.~A.~J., {van Ballegooijen}, A.~A., {Jardine}, M., {Collier
  Cameron}, A., Aug. 2002. {The Coronal Topology of the Rapidly Rotating K0
  Dwarf AB Doradus. I. Using Surface Magnetic Field Maps to Model the Structure
  of the Stellar Corona}. Astrophysical Journal. 575, 1078--1086.

\bibitem[{{Kasting}(1993)}]{kasting1993}
{Kasting}, J.~F., Feb. 1993. {Earth's early atmosphere}. Science 259, 920--926.

\bibitem[{{Keenan} et~al.(2008){Keenan}, {Jess}, {Aggarwal}, {Thomas},
  {Brosius}, and {Davila}}]{Keenan2008}
{Keenan}, F.~P., {Jess}, D.~B., {Aggarwal}, K.~M., {Thomas}, R.~J., {Brosius},
  J.~W., {Davila}, J.~M., Sep. 2008. {Emission lines of FeX in active region
  spectra obtained with the Solar Extreme-ultraviolet Research Telescope and
  Spectrograph}. Monthly Notices of the RAS. 389, 939--948.

\bibitem[{{Keenan} et~al.(2010){Keenan}, {Milligan}, {Jess}, {Aggarwal},
  {Mathioudakis}, {Thomas}, {Brosius}, and {Davila}}]{Keenan2010}
{Keenan}, F.~P., {Milligan}, R.~O., {Jess}, D.~B., {Aggarwal}, K.~M.,
  {Mathioudakis}, M., {Thomas}, R.~J., {Brosius}, J.~W., {Davila}, J.~M., May
  2010. {Emission lines of FeXI in the 257-407{\AA} wavelength region observed
  in solar spectra from EIS/Hinode and SERTS}. Monthly Notices of the RAS. 404,
  1617--1624.

\bibitem[{{Laming} et~al.(1995){Laming}, {Drake}, and {Widing}}]{Laming1995}
{Laming}, J.~M., {Drake}, J.~J., {Widing}, K.~G., Apr. 1995. {Stellar coronal
  abundances. 3: The solar first ionization potential effect determined from
  full-disk observation}. Astrophysical Journal. 443, 416--422.

\bibitem[{{Malinovsky} and {Heroux}(1973)}]{Malinovsky1973}
{Malinovsky}, L., {Heroux}, M., May 1973. {An Analysis of the Solar
  Extreme-Ultraviolet Between 50 and 300 A}. Astrophysical Journal. 181,
  1009--1030.

\bibitem[{{McComas} et~al.(2007){McComas}, {Velli}, {Lewis}, {Acton},
  {Balat-Pichelin}, {Bothmer}, {Dirling}, {Feldman}, {Gloeckler}, {Habbal},
  {Hassler}, {Mann}, {Matthaeus}, {McNutt}, {Mewaldt}, {Murphy}, {Ofman},
  {Sittler}, {Smith}, and {Zurbuchen}}]{mccomas2007}
{McComas}, D.~J., {Velli}, M., {Lewis}, W.~S., {Acton}, L.~W.,
  {Balat-Pichelin}, M., {Bothmer}, V., {Dirling}, R.~B., {Feldman}, W.~C.,
  {Gloeckler}, G., {Habbal}, S.~R., {Hassler}, D.~M., {Mann}, I., {Matthaeus},
  W.~H., {McNutt}, R.~L., {Mewaldt}, R.~A., {Murphy}, N., {Ofman}, L.,
  {Sittler}, E.~C., {Smith}, C.~W., {Zurbuchen}, T.~H., Mar. 2007.
  {Understanding coronal heating and solar wind acceleration: Case for in situ
  near-Sun measurements}. Reviews of Geophysics 45, 1004--+.

\bibitem[{McFadden and Merrill(2000)}]{mcfadden2000}
McFadden, P., Merrill, R., 2000. Evolution of the geomagnetic reversal rate
  since 160 ma: Is the process continuous? J. Geophys. Res. 105, 28,455Ð28,460.

\bibitem[{{Merrill} and {McElhinny}(1983)}]{merrill1983}
{Merrill}, R.~T., {McElhinny}, M.~W., 1983. {The earth's magnetic field. Its
  history, origin and planetary perspective.}

\bibitem[{{Mewe} et~al.(1995){Mewe}, {Kaastra}, {Schrijver}, {van den Oord},
  and {Alkemade}}]{mewe1995}
{Mewe}, R., {Kaastra}, J.~S., {Schrijver}, C.~J., {van den Oord}, G.~H.~J.,
  {Alkemade}, F.~J.~M., Apr. 1995. {EUV spectroscopy of cool stars. I. The
  corona of {$\alpha$} Centauri observed with EUVE. R}. \aap 296, 477--+.

\bibitem[{{Ness} et~al.(2004){Ness}, {G{\"u}del}, {Schmitt}, {Audard}, and
  {Telleschi}}]{Ness2004}
{Ness}, J., {G{\"u}del}, M., {Schmitt}, J.~H.~M.~M., {Audard}, M., {Telleschi},
  A., Nov. 2004. {On the sizes of stellar X-ray coronae}. Astron. and
  Astrophys. 427, 667--683.

\bibitem[{{Ness} et~al.(2002){Ness}, {Schmitt}, {Burwitz}, {Mewe}, {Raassen},
  {van der Meer}, {Predehl}, and {Brinkman}}]{ness2002}
{Ness}, J., {Schmitt}, J.~H.~M.~M., {Burwitz}, V., {Mewe}, R., {Raassen},
  A.~J.~J., {van der Meer}, R.~L.~J., {Predehl}, P., {Brinkman}, A.~C., Nov.
  2002. {Coronal density diagnostics with Helium-like triplets: CHANDRA-LETGS
  observations of Algol, Capella, Procyon, epsilon Eri, alpha Cen A{\&}B, UX
  Ari, AD Leo, YY Gem, and HR 1099}. \aap 394, 911--926.

\bibitem[{{Petit} et~al.(2008){Petit}, {Dintrans}, {Solanki}, {Donati},
  {Auri{\`e}re}, {Ligni{\`e}res}, {Morin}, {Paletou}, {Ramirez Velez},
  {Catala}, and {Fares}}]{petit2008}
{Petit}, P., {Dintrans}, B., {Solanki}, S.~K., {Donati}, J., {Auri{\`e}re}, M.,
  {Ligni{\`e}res}, F., {Morin}, J., {Paletou}, F., {Ramirez Velez}, J.,
  {Catala}, C., {Fares}, R., Jul. 2008. {Toroidal versus poloidal magnetic
  fields in Sun-like stars: a rotation threshold}. Monthly Notices of the RAS.
  388, 80--88.

\bibitem[{{Pevtsov} et~al.(2003){Pevtsov}, {Fisher}, {Acton}, {Longcope},
  {Johns-Krull}, {Kankelborg}, and {Metcalf}}]{pevtsov2003}
{Pevtsov}, A.~A., {Fisher}, G.~H., {Acton}, L.~W., {Longcope}, D.~W.,
  {Johns-Krull}, C.~M., {Kankelborg}, C.~C., {Metcalf}, T.~R., Dec. 2003. {The
  Relationship Between X-Ray Radiance and Magnetic Flux}. Astrophysical
  Journal. 598, 1387--1391.

\bibitem[{Powell~et al.(1999)}]{powell1999}
Powell~et al., K., 1999. A solution-adaptive upwind scheme for ideal
  magnetohydrodynamics. Journal of Computational Physics 154, 284--309.

\bibitem[{{Ribas} et~al.(2005){Ribas}, {Guinan}, {G{\"u}del}, and
  {Audard}}]{ribas2005}
{Ribas}, I., {Guinan}, E.~F., {G{\"u}del}, M., {Audard}, M., Mar. 2005.
  {Evolution of the Solar Activity over Time and Effects on Planetary
  Atmospheres. I. High-Energy Irradiances (1-1700 {\AA})}. Astrophysical
  Journal. 622, 680--694.

\bibitem[{{Ribas} et~al.(2010){Ribas}, {Porto de Mello}, {Ferreira},
  {H{\'e}brard}, {Selsis}, {Catal{\'a}n}, {Garc{\'e}s}, {do Nascimento}, and
  {de Medeiros}}]{ribas2010}
{Ribas}, I., {Porto de Mello}, G.~F., {Ferreira}, L.~D., {H{\'e}brard}, E.,
  {Selsis}, F., {Catal{\'a}n}, S., {Garc{\'e}s}, A., {do Nascimento}, J.~D.,
  {de Medeiros}, J.~R., May 2010. {Evolution of the Solar Activity Over Time
  and Effects on Planetary Atmospheres. II. $\kappa^{1}$ Ceti, an Analog of the
  Sun when Life Arose on Earth}. Astrophysical Journal. 714, 384--395.

\bibitem[{Ridley(2005)}]{ridley2005}
Ridley, A.~J., 2005. A new formulation for the ionospheric cross polar cap
  potential including saturation effects. Annales Geophysicae 23~(11),
  3533--3547.
\newline\urlprefix\url{http://www.ann-geophys.net/23/3533/2005/}

\bibitem[{{Ridley} and {Liemohn}(2002)}]{ridley2002}
{Ridley}, A.~J., {Liemohn}, M.~W., Aug. 2002. {A model-derived storm time
  asymmetric ring current driven electric field description}. Journal of
  Geophysical Research (Space Physics) 107, 1151--+.

\bibitem[{{Rosing} et~al.(2010){Rosing}, {Bird}, {Sleep}, and
  {Bjerrum}}]{rosing2010}
{Rosing}, M.~T., {Bird}, D.~K., {Sleep}, N.~H., {Bjerrum}, C.~J., Apr. 2010.
  {No climate paradox under the faint early Sun}. Nature 464, 744--747.

\bibitem[{{Roussev} et~al.(2003){Roussev}, {Gombosi}, {Sokolov}, {Velli},
  {Manchester}, {DeZeeuw}, {Liewer}, {T{\'o}th}, and {Luhmann}}]{roussev2003}
{Roussev}, I.~I., {Gombosi}, T.~I., {Sokolov}, I.~V., {Velli}, M.,
  {Manchester}, IV, W., {DeZeeuw}, D.~L., {Liewer}, P., {T{\'o}th}, G.,
  {Luhmann}, J., Sep. 2003. {A Three-dimensional Model of the Solar Wind
  Incorporating Solar Magnetogram Observations}. Astrophysical Journal,
  Letters. 595, L57--L61.

\bibitem[{{Sagan} and {Mullen}(1972)}]{sagan1972}
{Sagan}, C., {Mullen}, G., Jul. 1972. {Earth and Mars: Evolution of Atmospheres
  and Surface Temperatures}. Science 177, 52--56.

\bibitem[{Saito et~al.(1978)Saito, Sakurai, and Yumoto}]{saito1978}
Saito, T., Sakurai, T., Yumoto, K., 1978. The {E}arth's paleomagnetosphere as
  the third type of planetary magnetosphere. Planetary and Space Science 26,
  413--422.

\bibitem[{{Schmitt} et~al.(1996){Schmitt}, {Drake}, {Haisch}, and
  {Stern}}]{schmitt1996}
{Schmitt}, J.~H.~M.~M., {Drake}, J.~J., {Haisch}, B.~M., {Stern}, R.~A., Aug.
  1996. {A Close Look at the Coronal Density of Procyon}. \apj 467, 841--+.

\bibitem[{{Shestov} et~al.(2009){Shestov}, {Urnov}, {Kuzin}, {Zhitnik}, and
  {Bogachev}}]{Shestov2009}
{Shestov}, S.~V., {Urnov}, A.~M., {Kuzin}, S.~V., {Zhitnik}, I.~A., {Bogachev},
  S.~A., Jan. 2009. {Electron density diagnostics for various plasma structures
  of the solar corona based on Fe XI-FeXIII lines in the range 176 207 {\AA}
  measured in the SPIRIT/CORONAS-F experiment}. Astronomy Letters 35, 45--56.

\bibitem[{{Siscoe} and {Chen}(1975)}]{siscoe1975}
{Siscoe}, G.~L., {Chen}, C., Dec. 1975. {The paleomagnetosphere}. J. Geophys.
  Res. 80, 4675--4680.

\bibitem[{{Skumanich}(1972)}]{skumanich1972}
{Skumanich}, A., Feb. 1972. {Time Scales for CA II Emission Decay, Rotational
  Braking, and Lithium Depletion}. Astrophysical Journal. 171, 565--+.

\bibitem[{Stevenson(2003)}]{stevenson2003}
Stevenson, D., 2003. Planetary magnetic fields. Earth and Planetary Sciences
  208, 1--11.

\bibitem[{Tarduno et~al.(2010)Tarduno, Cottrell, Watkeys, Hofmann, Doubrovine,
  Mamajek, Liu, Sibeck, Neukirch, and Usui}]{tarduno2010}
Tarduno, J.~A., Cottrell, R.~D., Watkeys, M.~K., Hofmann, A., Doubrovine,
  P.~V., Mamajek, E.~E., Liu, D., Sibeck, D.~G., Neukirch, L.~P., Usui, Y.,
  2010. {Geodynamo, Solar Wind, and Magnetopause 3.4 to 3.45 Billion Years
  Ago}. Science 327~(5970), 1238--1240.

\bibitem[{{Testa} et~al.(2004){Testa}, {Drake}, and {Peres}}]{Testa2004}
{Testa}, P., {Drake}, J.~J., {Peres}, G., Dec. 2004. {The Density of Coronal
  Plasma in Active Stellar Coronae}. Astrophysical Journal. 617, 508--530.

\bibitem[{Toth~et al(2005)}]{toth2005}
Toth~et al, G., 2005. Space {W}eather {M}odeling {F}ramework: {A} new tool for
  the space science community. J. Geophys. Res. 110.

\bibitem[{{Totten} et~al.(1995){Totten}, {Freeman}, and {Arya}}]{totten1995}
{Totten}, T.~L., {Freeman}, J.~W., {Arya}, S., Jan. 1995. {An empirical
  determination of the polytropic index for the free-streaming solar wind using
  HELIOS 1 data}. J. Geophys. Res. 100, 13--17.

\bibitem[{Vogt and Glassmeier(2001)}]{vogt2001}
Vogt, J., Glassmeier, K.-H., 2001. Modelling the paleomagnetosphere: Strategy
  and first results. Advances in Space Research 28, 863--868.

\bibitem[{{Vogt} et~al.(2004){Vogt}, {Zieger}, {Stadelmann}, {Glassmeier},
  {Gombosi}, {Hansen}, and {Ridley}}]{vogt2004}
{Vogt}, J., {Zieger}, B., {Stadelmann}, A., {Glassmeier}, K., {Gombosi}, T.~I.,
  {Hansen}, K.~C., {Ridley}, A.~J., Dec. 2004. {MHD simulations of quadrupolar
  paleomagnetospheres}. JGR (Space Physics) 109~(A18), 12221--+.

\bibitem[{{Wang} and {Sheeley}(1990)}]{wang1990}
{Wang}, Y., {Sheeley}, Jr., N.~R., Jun. 1990. {Solar wind speed and coronal
  flux-tube expansion}. Astrophysical Journal. 355, 726--732.

\bibitem[{{Wood} et~al.(2002){Wood}, {M{\"u}ller}, {Zank}, and
  {Linsky}}]{wood2002}
{Wood}, B.~E., {M{\"u}ller}, H., {Zank}, G.~P., {Linsky}, J.~L., Jul. 2002.
  {Measured Mass-Loss Rates of Solar-like Stars as a Function of Age and
  Activity}. Astrophysical Journal. 574, 412--425.

\bibitem[{{Wood} et~al.(2005){Wood}, {M{\"u}ller}, {Zank}, {Linsky}, and
  {Redfield}}]{wood2005}
{Wood}, B.~E., {M{\"u}ller}, H., {Zank}, G.~P., {Linsky}, J.~L., {Redfield},
  S., Aug. 2005. {New Mass-Loss Measurements from Astrospheric Ly{$\alpha$}
  Absorption}. Astrophysical Journal, Letters. 628, L143--L146.

\bibitem[{{Zieger} et~al.(2006{\natexlab{a}}){Zieger}, {Vogt}, and
  {Glassmeier}}]{zieger2006}
{Zieger}, B., {Vogt}, J., {Glassmeier}, K., Jun. 2006{\natexlab{a}}. {Scaling
  relations in the paleomagnetosphere derived from MHD simulations}. JGR
  111~(A10), 6203--+.

\bibitem[{{Zieger} et~al.(2004){Zieger}, {Vogt}, {Glassmeier}, and
  {Gombosi}}]{zieger2004}
{Zieger}, B., {Vogt}, J., {Glassmeier}, K., {Gombosi}, T.~I., Jul. 2004.
  {Magnetohydrodynamic simulation of an equatorial dipolar paleomagnetosphere}.
  J. Geophys. Res. 109~(A18), 7205--+.

\bibitem[{{Zieger} et~al.(2006{\natexlab{b}}){Zieger}, {Vogt}, {Ridley}, and
  {Glassmeier}}]{zieger2006b}
{Zieger}, B., {Vogt}, J., {Ridley}, A.~J., {Glassmeier}, K.,
  2006{\natexlab{b}}. {A parametric study of magnetosphere ionosphere coupling
  in the paleomagnetosphere}. Advances in Space Research 38, 1707--1712.

\end{thebibliography}

%
%
%
%
%
%
%
%


%
%

\end{article}




\newpage
\clearpage

\begin{table}[htdp]
\begin{center}\small
\begin{tabular}{l|c|c|c|c|c}
\hline     
solar wind &  $u_{av}$  &  $n_{av}$ &    $B_{av}$  &   $T_{av}$ & mass flux\\
solution & (km/s) & (cm$^{-3}$) & (nT) & (K) & ($10^{-14}$ M$_{\odot}$/yr)\\
 \hline
 sf2df2 &   392.06   &     18.54    &     3.10  &   50542.81 & 5.55\\
sf2df5  &    413.26    &    20.41    &     3.44   &  53117.51 & 5.43\\
sf2df10  &     450.05   &     20.12  &       4.03  &   60625.57 & 5.01\\
sf2df10$\_$15d$\_$hd & 280.43   &    441.17 &       11.90 &    34261.75 & 109.50 \\
\hdashline
sf5df2     &  579.19   &     16.72   &      6.13  &   88623.79 & 6.30\\
sf10df2    &  388.56   &     27.08   &     7.86   & 49528.60 & 7.58\\
sf10df2$\_$15d &     381.95   &     27.05   &     9.20 &   50009.18 & 7.59\\
sf10df2$\_$15d$\_$hd & 266.00 &      547.99   &     24.76  &   32020.56   & 142.06 \\
\hdashline
shift00      &  444.57     &   10.67    &     2.41   &  62499.35 & 4.28\\
shift00$\_$up   & 448.66   &     10.44    &     2.54   &  63477.22 & -\\
shift00$\_$down &  440.13   &     10.81 &        2.29 &    60785.75 & -\\
shift00$\_$hd  & 287.34    &   373.98  &       8.01  &  34732.92 & 103.57\\
\hdashline
shift30    &   394.21    &    14.66  &       2.94   &  50702.75 & 4.93\\
shift60    &   392.06    &    18.54  &       3.10   &  50542.81 & 5.55\\
cr2074-shift00     &  307.59    &    22.31   &      2.01  &   39500.31 & 3.24\\
\hline
\end{tabular}
\end{center}

\caption{Average solar wind parameters for different solar wind solutions extracted along the trajectory of the ACE spacecraft, which is situated at the L1 Earth-Sun Lagrange point, at $\sim1$ AU. The mass flux decreases for stronger dipole magnetic field and increases for stronger sunspot magnetic field since the stronger spots close down open flux. Hence the fast wind is reduced and the 
corona is dominated by slow wind, which has a higher density, which seems to dominate the overall behavior of the mass flux. We did not calculate the mass flux for the "up" and "down" solutions. "hd" refers to a coronal base density of $10^{10}$ cm$^{-3}$ instead of the regular $5\times10^8$ cm$^{-3}$.}
\label{tab:massflux}
\end{table}

\begin{table}[htdp]
\begin{center}\small
\begin{tabular}{l|c|c|c|c|c||c|c|c|c|c}
\hline 
& \multicolumn{5}{|c||}{Input parameters} & \multicolumn{5}{|c}{Output parameters}\\
\hline
Scenario & $u$ & $n$ & $b_z$ & $B$ & tilt & $R_{mp}$ & $R_{dd}$ & $R_{ns}$ & $\theta_{pc}$ & $A_{pc}$  \\ 
& (km/s) & (cm$^{-3}$) & (nT) & ($10^4$ nT) & ($^{\circ}$)& ($R_E$) & ($R_E$) & ($R_E$) & ($^{\circ}$) &   \\ 
\hline
A & 380 & 27 & -10 & 4 & 0  & 8.5 & 11.5 & 13.5  & 18/24& 1.05  \\
B & 380 & 27 & -10 & 1.5 & 0 & 5.75 & 8 & 10  & 24/28 & 1.27  \\
C & 380 & 27 & -10 & 0.75 & 0 & 4.25 & 6 & 8.2   &  29/30 &1.46  \\
\hdashline
D & 460 & 18 & -4 & 4 & 0 & 8.5 & 11 & 13.2  & 16/24 &0.98  \\
E & 460 &  18 & -4 & 4& 45 & 8.75& 10  & 15$^a$   & 12/23 &0.91 \\
F & 460 & 18 & -4 & 4 & 90 & 8.5 & 9.8& 9.25$^b$    & 6/20 & 0.77 \\
\hdashline
G & 460 & 18 & -4 & 1.5 & 0 & 6 & 9 & 10    &20/27  & 1.18 \\

H & 460 & 18 & +4 & 1.5 & 0 &6.25 & 8.25 & 9   & -  & -  \\
\hdashline
I & 285 & 375 & -8 & 1.5 & 0 &4.25 & 6.13 &  7  & 23/31 & 1.38  \\
J& 280 & 440 & -12 &1.5 & 0 & 4.4 & 5.8 & 7 & 25/31 & 1.42 \\
K& 280 & 440 & -12 &4 & 0 & 6 & 8.6 & 9.3 & 19/28 & 1.17 \\
L& 265 & 545 & -25 & 1.5 & 0 & 4 & 5 & 7 & 29/32 & 1.52 \\
M&265 & 545 & -25 & 4 & 0 & 5.75 & 8 & 9.3 & 22/28 & 1.26 \\ 
\hdashline
N & 435 & 5.5 & -6 & 3 & 11 &9.5 &12.75 & 15.75$^c$    &17/23$^d$& 1  \\
\hline
\end{tabular}
\end{center}
\caption{Input and output parameters for the solar wind/paleomagnetosphere scenarios used in the GM MHD simulations. Ouput parameters are: Magnetopause subsolar stand-off distance; Magnetopause dawn-dusk flank distance; Magnetopause north-south flank distance; Polar cap magnetic co-latitude (dayside / nightside) ; Polar cap area normalized by present Earth scenario N.\\ $^a$: North flank distance. South flank distance is 11.25 $R_E$. $^b$: North flank distance. South flank distance is 11 $R_E$. $^c$: North flank distance. South flank distance is 14.25 $R_E$. $^d$: Present value of polar cap co-latitude: 18.1$^\circ$ \citep{tarduno2010}}
\label{tab:gmcases}
\end{table}

%
%
%


\end{document}